\documentclass[utf8]{template_2022/frontiersinFPHY_FAMS} 

\pdfoutput=1
\usepackage{url,hyperref,lineno,microtype,subcaption}
\usepackage[onehalfspacing]{setspace}
\usepackage{listings}
\usepackage{braket}
\usepackage[ISO]{diffcoeff}
\usepackage{xspace}
\usepackage{dsfont}
\usepackage{bm}
\usepackage{isotope}

\usepackage{cellspace}
\usepackage{booktabs}
\usepackage{comment}

\newcommand{\trial}{\widetilde}
\newcommand{\subspace}{\widetilde}
\newcommand{\coeff}{\beta}
\newcommand{\coeffs}{\vec{\coeff}}
\newcommand{\coeffsopt}{\vec{\beta}_{\star}}

\newcommand{\eg}{\textit{e.g.}\xspace}
\newcommand{\ie}{\textit{i.e.}}

\newcommand{\ritzbasis}{X}

\newcommand{\action}{\mathcal{S}}
\newcommand{\nbasis}{n_b}
\newcommand{\nsimulator}{N_h}

\newcommand{\lagmult}{\lambda_\star}

\newcommand{\dU}{\Delta \widetilde U}
\newcommand{\param}{\theta}
\newcommand{\params}{\boldsymbol{\theta}}
\newcommand{\umatrix}{\boldsymbol{u}}
\newcommand{\genkvp}{\mathcal{L}}

\newcommand{\identity}{\mathds{1}}
\newcommand{\trialfunc}{\xi}
\newcommand{\testfunc}{\zeta}

\newcommand{\new}[1]{#1}

\newcommand{\trans}{\intercal}

\def\diffd{\mathrm{d}}  

\DeclareDocumentCommand\differential{ o g d() }{ 
    \IfNoValueTF{#2}{
        \IfNoValueTF{#3}
            {\diffd\IfNoValueTF{#1}{}{^{#1}}}
            {\mathinner{\diffd\IfNoValueTF{#1}{}{^{#1}}\argopen(#3\argclose)}}
        }
        {\mathinner{\diffd\IfNoValueTF{#1}{}{^{#1}}#2} \IfNoValueTF{#3}{}{(#3)}}
    }
\DeclareDocumentCommand\dd{}{\differential} 

\graphicspath{{figures/}{template_2022/}}

\definecolor{linkcolor}{rgb}{0,0,0.80} 
\hypersetup{%
    pdfsubject=Paper,
    pdfkeywords={nuclear physics} {Bayesian} {chiral EFT} {emulator} {reduced-order models} {variational principles},
    unicode = true,
    breaklinks = true,
    colorlinks = true,
    linkcolor = linkcolor,
    citecolor = linkcolor,
    menucolor = linkcolor,
    urlcolor = linkcolor
}

\def\keyFont{\fontsize{8}{11}\helveticabold }
\def\firstAuthorLast{Drischler, Melendez, Furnstahl, Garcia, and Zhang} 
\def\Authors{C. Drischler,$^{1,2,*}$ 
J.~A. Melendez,$^{3}$ 
R.~J. Furnstahl,$^{3}$ 
A.~J. Garcia,$^{3}$ 
and Xilin Zhang$^{2}$}
\def\Address{$^{1}$Department of Physics and Astronomy  \& Institute of Nuclear and Particle Physics, Ohio University, Athens, OH 45701, USA \\ $^{2}$Facility for Rare Isotope Beams, Michigan State University, MI 48824, USA \\
$^{3}$Department of Physics, The Ohio State University, Columbus, OH 43210, USA  }
\def\corrAuthor{Christian Drischler}
\def\corrEmail{drischler@ohio.edu}

\begin{document}
\firstpage{1}

\title[BUQEYE Guide to Projection-Based Emulators...]{BUQEYE Guide to Projection-Based Emulators in Nuclear Physics}

\author[\firstAuthorLast]{\Authors} 
\address{} 
\correspondance{} 

\extraAuth{}
\twocolumn[
\begin{@twocolumnfalse}
\maketitle

\begin{abstract}  


\noindent
The BUQEYE collaboration (Bayesian Uncertainty Quantification: Errors in Your EFT) presents a pedagogical introduction 
to projection-based, reduced-order emulators for applications in low-energy nuclear physics.
The term \emph{emulator} refers here to a fast surrogate model 
capable of reliably approximating high-fidelity models.
As the general tools employed by these emulators are not yet well-known in the nuclear physics community, we discuss variational and Galerkin projection methods, emphasize the benefits of offline-online decompositions, and explore how these concepts lead to emulators for bound and scattering systems that enable fast \& accurate calculations using many different model parameter sets.
We also point to future extensions and applications of these emulators for nuclear physics, guided by the mature field of model (order) reduction.
All examples discussed here and more are available as interactive, open-source Python code so that practitioners 
can readily adapt projection-based emulators for their own work.

\tiny
 \keyFont{ \section{Keywords:} 
 emulators, 
 reduced-order models,
 model order reduction,
 nuclear scattering, 
 uncertainty quantification, 
 effective field theory, 
 variational principles,
 Galerkin projection} 
\end{abstract}
\end{@twocolumnfalse}
]

\section{Introduction}

Nuclear systems are notoriously complex. But typically, our theoretical modeling of nuclear phenomena 
contains superfluous information
for quantities of interest.
Model order reduction (MOR) refers to powerful techniques that enable us to reduce a system's complexity systematically (\eg, see References~\cite{Benner_2017aa,Benner2017modelRedApprox,benner2015survey} for comprehensive introductions).
These techniques enable emulators, 
which are low-dimensional surrogate models capable of rapidly and reliably approximating high-fidelity models,
making practical otherwise impractical calculations.
But the nuclear physics community has barely scratched the surface of the types of emulators that could be crafted or explored their full range of applications.

A fertile area for new emulators is uncertainty quantification (UQ)~\cite{Zhang:2015ajn, Neufcourt:2019qvd,King:2019sax,Ekstrom:2019lss,Catacora-Rios:2020xgx,Wesolowski:2021cni,Svensson:2021lzs,Odell:2021tqd,Djarv:2021pjc,Alnamlah:2022eae} in nuclear physics, which is the general theme of this Frontiers Research Topic~\cite{ResearchTopicUQ}.
\new{Quantifying theoretical uncertainties rigorously is crucial for comparing theory predictions with experimental and/or observational constraints and performing model comparison and/or mixing~\cite{Phillips:2020dmw}.
However, UQ has only recently drawn much attention as nuclear theory has entered the precision era.}
Bayesian parameter estimation for nuclear effective field theory (EFT) and optical models, UQ for nuclear structure pushing toward larger masses and for reactions across the chart of nuclides, experimental design~\cite{Melendez:2020ikd,Phillips:2020dmw,Farr:2021fyc} for the next generation of precision experiments probing the nuclear dripline,
and many other applications will all benefit from emulators.
\new{This Research Topic~\cite{ResearchTopicUQ} already contains several new applications of emulators for nuclear physics.}
Key to the wider adoption of these tools is the evangelization of their potential and the creation of pedagogical guides for those first starting in this field~\cite{Melendez:2022kid}.
This article is aimed at 
both goals.


To do so, the BUQEYE collaboration (Bayesian Uncertainty Quantification: Errors in Your EFT)~\cite{BUQEYEweb} has created a rather unconventional document comprised of the article you are reading now along with a companion website~\cite{companionwebsite} containing interactive supplemental material and source code that generates all the results shown, and much more.
Interested individuals can dynamically generate different versions of this document based on tunable parameters.
We hope that this format encourages readers to experiment and build upon the examples presented here, thereby facilitating new applications.


Various types of 
emulators have already been applied with success within nuclear physics.
A non-exhaustive list of applications includes References~\cite{Higdon:2014tva,Frame:2017fah,Sarkar:2020mad,Sarkar:2021fpz,Konig:2019adq,Demol:2019yjt,Ekstrom:2019lss,Bai:2021xok,Demol:2020mzd,Yoshida:2021jbl,Wesolowski:2021cni,Furnstahl:2020abp,Melendez:2021lyq,Drischler:2021qoy,Zhang:2021jmi,Drischler:2022yfb,Anderson:2022jhq,Giuliani:2022yna,Surer:2022lhs,Bai:2022hjg,Kravvaris:2020lhp,Yapa:2022nnv,Francis:2022zib,Zare:2022cdw}.
But as emphasized in References~\cite{Melendez:2022kid,Bonilla:2022rph}, there is a broad and relatively mature MOR literature outside of nuclear physics waiting to be exploited (\eg, see Reference~\cite{Benner2020Volume1DataDriven} for an overview of the universe of MOR approaches).
Our goal in this guide will be to facilitate this exploitation through a selective treatment of physics-informed, projection-based emulators relevant to a wide range of nuclear physics problems.

To this end, we organize this guide as follows. Section~\ref{sec:eigen-emulators} focuses on emulators for bound-state calculations using subspace-projection methods. 
We then provide a more general introduction to MOR for solving differential equations in Section~\ref{sec:model-reduction}, which leads to our discussion of scattering emulators in Section~\ref{sec:scattering-emulators}.
Section~\ref{sec:conclusion} concludes with a summary and outlook.
Throughout, we draw connections between variational and Galerkin projection methods and illustrate these concepts with pedagogical examples, supplemented by source code on the companion website~\cite{companionwebsite}.

\section{Eigen-Emulators} \label{sec:eigen-emulators}

In this section, we discuss the construction of fast \& accurate emulators for bound-state calculations.
Given a (Hermitian) Hamiltonian $H(\params)$, we aim to find the solutions $\{E(\params), \ket{\psi(\params) }\}$ of the parametric Schrödinger Equation
\begin{equation} \label{eq:generic_eigenvalue_problem}
    H(\params)\ket{\psi(\params)} = E(\params) \ket{\psi(\params)} ,
\end{equation}
subject to the normalization $\braket{\psi(\params)|\psi(\params)} = 1$, in the parameter space $\params$.
The components of the vector $\params$ may be model parameters, such as the low-energy couplings of a nuclear EFT, or other parameters describing the system of interest~\cite{Zhang:2021jmi,Yapa:2022nnv}. 
We consider here cases in which Equation~\eqref{eq:generic_eigenvalue_problem} can be solved with high fidelity, but doing so requires a significant amount of computing time, \eg, repeated calculations in the parameter space $\params$ for Monte Carlo sampling or optimization tasks are computationally demanding or even prohibitively slow.
In the following, we will discuss how the Ritz variational principle and the Galerkin method can be used to construct rapid and reliable\footnote{A \emph{reliable} emulator may not necessarily be required to be highly accurate, \eg, if the other uncertainties of the theoretical calculation dominate the overall uncertainty budget.} emulators that facilitate these calculations. 

\subsection{Variational approach}
\label{sec:eigen-emulators_variational}

To construct an emulator for bound state calculations, we use here the Rayleigh--Ritz method\footnote{%
For a critical commentary on the history of the method's name, see, \eg, References~\cite{LEISSA2005961,ILANKO2009731}.%
} and thus consider the energy functional
\begin{equation} \label{eq:ritz_functional}
    \mathcal{E}[\trial\psi] = \braket{\trial\psi | H(\params) | \trial\psi} - \subspace{E}(\params) (\braket{\trial\psi | \trial\psi} - 1),
\end{equation}
%
where the Lagrange multiplier $\subspace{E}(\params)$ (also known as Ritz value)
imposes the normalization condition $\braket{\trial\psi | \trial\psi} = 1$ for bound states. The General Ritz Theorem~\cite{Suzuki:1998bn}\footnote{Many helpful theorems relevant to the Rayleigh--Ritz method can be found in Section~3 in Reference~\cite{Suzuki:1998bn}.} states that the functional~\eqref{eq:ritz_functional} is stationary about all (discrete) solutions of the Schrödinger Equation~\eqref{eq:generic_eigenvalue_problem}, not just the ground state solution, which can be seen by 
imposing the stationary condition 
\begin{align} \label{eq:stationarity_condition}
    \delta\mathcal{E}[\trial\psi] \equiv 0 
    &= 2\braket{\delta\trial\psi | [H(\params)-\subspace{E}(\params)] | \trial\psi} \notag \\ 
    &\quad \null - \delta\subspace{E}(\params) [{\braket{\trial\psi | \trial\psi}} - 1] ,
\end{align}
and noting that Equation~\eqref{eq:stationarity_condition} is only fulfilled for arbitrary variations $\bra{\delta\trial\psi}$ if $\ket{\trial\psi}$ is a solution of the Schrödinger Equation~\eqref{eq:generic_eigenvalue_problem} with $\trial E(\params) = E(\params)$.

Let us now define the trial wave function we use in conjunction with the functional~\eqref{eq:ritz_functional}:
\begin{subequations} \label{eq:trial_general}
 \begin{align}
    \ket{\trial\psi} &= \sum_{i=1}^{\nbasis} \coeff_i\ket{\psi_i} \equiv X\coeffs, \\
    X & =
    \bgroup
    \renewcommand*{\arraystretch}{1.1}
    \begin{bmatrix}
        \ket{\psi_1} & \ket{\psi_2} & \cdots &  \ket{\psi_{\nbasis}}
    \end{bmatrix},
    \egroup
 \end{align}
\end{subequations}
where the column-vector $\coeffs$ contains the to-be-determined coefficients and the row-vector%
\footnote{\new{In a representation of $H$, the $\psi_i$ corresponding to $\ket{\psi_i}$ are the $\nbasis$ columns of the matrix $X$ in that representation.
}} $X$ the (in principle) arbitrary basis states.
Here, we use \emph{snapshots} of high-fidelity solutions of the Schrödinger Equation~\eqref{eq:generic_eigenvalue_problem} at a set of given parameter values; \ie, $\{\ket{\psi_i} \equiv \ket{\psi(\params_i)}\}_{i=1}^{\nbasis}$~\cite{Benner2017modelRedApprox,Benner20201,Buchan2013EigenvaluePOD,Quarteroni:218966}.
No assumption has been made as to how to obtain the high-fidelity solutions.

\begin{figure}[tb]
    \centering
    \includegraphics[width=\linewidth]{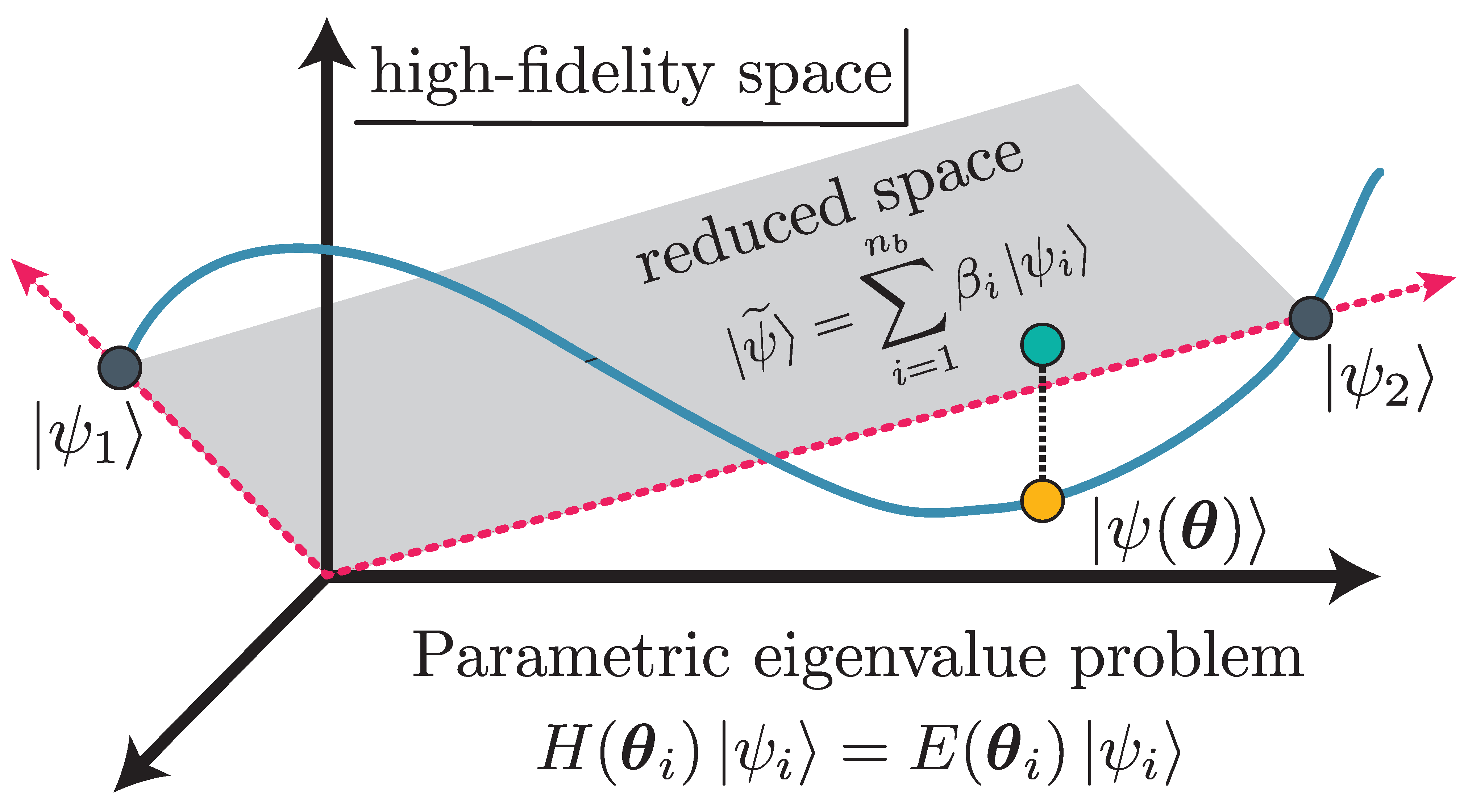}
    \caption{Illustration of a projection-based emulator using only two snapshots $\ket{\psi_i} \equiv \ket{\psi(\params_i)}$ (dark gray points). 
    These snapshots are high-fidelity solutions of the Schrödinger Equation~\eqref{eq:generic_eigenvalue_problem}, which
    span the subspace of the reduced-order model, as indicated by the red arrows and the gray plane.
    The trajectory of a high-fidelity eigenvector is denoted by the blue curve.
    The orange dot depicts an eigenvector $\ket{\psi(\params)}$ along the trajectory that,
    when projected onto the reduced space, corresponds to the turquoise point; hence, the difference between the orange and turquoise points represents the error due to the emulator's subspace projection (\ie, the dotted line).
    Inspired by Figure~2.1 in Reference~\cite{Benner2017modelRedApprox}.
    }
    \label{fig:illustration_rbm}
\end{figure}

Figure~\ref{fig:illustration_rbm} motivates the efficacy of snapshot-based trial functions.
Although a given eigenvector $\ket{\psi(\params)}$ obtained from a high-fidelity solver resides in a high-dimensional (or even infinite-dimensional) space, the trajectory traced out by continuous variations in $\params$ remains in a relatively low-dimensional subspace (as illustrated by the gray plane).
Hence, linear combinations of high-fidelity eigenvectors spanning this subspace (\ie, the snapshots) make extremely effective trial wave functions for variational calculations.
In nuclear physics, snapshot-based emulators already have accurately approximated ground-state properties, such as binding energies, charge radii~\cite{Konig:2019adq, Ekstrom:2019lss, Wesolowski:2021cni}, and transition matrix elements~\cite{Wesolowski:2021cni,Yoshida:2021jbl}, and have been explored for applications to excited states~\cite{Franzke:2021ofs}.


Given the trial wave function~\eqref{eq:trial_general}, we determine the coefficients $\coeffs_\star$ that render $\mathcal{E}[\trial \psi = X\coeffs]$ stationary under variations $\ket{\delta \trial \psi} = X\ket{\delta \coeffs}$ of the trial wave function, as opposed to arbitrary variations.
Solving for the optimal $\coeffs_\star$ occurs then in the low-dimensional space spanned by the basis elements in $X$ (\ie, the red arrows in Figure~\ref{fig:illustration_rbm}) rather than in the high-dimensional space in which $\ket{\psi}$ resides.
From the stationarity condition~\eqref{eq:stationarity_condition}, we obtain the reduced-order model~\cite{Melendez:2020xcs}
\begin{subequations} \label{eq:ritz_condition}
    \begin{align}
    \subspace{H}(\params)\coeffsopt(\params) & = \subspace{E}(\params)\subspace{N} \coeffsopt(\params) , \label{eq:ritz_gen_eigvp}\\
    \coeffsopt^\dagger(\params) \subspace{N} \coeffsopt(\params) & = 1 ,
\end{align}
\end{subequations}
where $\subspace{H}(\params) \equiv X^\dagger H(\params) X$ is the subspace-projected Hamiltonian and $\subspace{N} \equiv X^\dagger X$ the norm matrix in the snapshot basis.
\new{As opposed to $H(\params)$ in Equation~\eqref{eq:generic_eigenvalue_problem}, $\subspace{H}(\params)$ (and likewise $\subspace{N}$) is a $\nbasis \times \nbasis$ Hermitian matrix,}
\begin{equation}
  \subspace{H}(\params) = 
    \begin{bmatrix}
      \braket{\psi_1|H(\params)|\psi_1} & \cdots & \braket{\psi_1|H(\params)|\psi_{\nbasis}}\\
        \vdots & \ddots &  \vdots \\  
        \braket{\psi_{\nbasis}|H(\params)|\psi_1} & \cdots & \braket{\psi_{\nbasis}|H(\params)|\psi_{\nbasis}}
    \end{bmatrix} .
\end{equation}

\new{In practice, the generalized eigenvalue problem~\eqref{eq:ritz_condition} may experience numerical instabilities due to small singular values in $\subspace{N}$.
(The instabilities also appear in reduced-order modeling of differential equations, see Section~\ref{sec:model-reduction}.)
One way to ameliorate these instabilities 
is to orthonormalize the snapshots in $X$, hence yielding $\subspace{N} = \identity$.
This approach could also permit some efficiency gains, by excluding the least important vectors as measured by their singular values.
Alternatively, Reference~\cite{Hicks:2022ovs} recently introduced a trimmed sampling algorithm that can substantially reduce the effects of noise in solving generalized eigenvalue problems.
Finally, a well-known approach to regularize both generalized eigenvalue problems and matrix inversion is through the use of a nugget~\cite{Neumaier98solvingill-conditioned,engl1996regularization}.
Here, a regularization parameter $\nu\ll 1$ (called a nugget) is added to the diagonal of the ill-conditioned matrix one wishes to invert (here, $\subspace{N}$), thus shifting its singular values.}

By solving the generalized eigenvalue problem~\eqref{eq:ritz_condition},
one obtains $\nbasis$ pairs $\{\subspace{E}(\params),\coeffsopt(\params)\}$ consisting of a Lagrange multiplier (\ie, an eigenvalue) and its corresponding coefficient vector.
Let us index the eigenvalues of the emulator equation~\eqref{eq:ritz_condition} and Schr{\"o}dinger Equation~\eqref{eq:generic_eigenvalue_problem} in ascending order; that is, $\subspace{E}_n \leqslant \subspace{E}_{n+1}$ and ${E}_n \leqslant {E}_{n+1}$, respectively, with $n=1$ indicating the lowest eigenvalue.
If the snapshot basis $X$ in the trial wave function~\eqref{eq:trial_general} contains $\nbasis$ linearly independent states,
then the Min--Max Theorem~\cite{Suzuki:1998bn} asserts that each Lagrange multiplier,
\begin{equation} \label{eq:variational_bounds}
    \subspace{E}_n(\params) \geqslant E_n(\params) \quad \text{for $1 \leqslant n \leqslant \nbasis$},
\end{equation}
provides an upper bound on its corresponding eigenvalue of the Schr{\"o}dinger Equation~\eqref{eq:generic_eigenvalue_problem}.\footnote{For non-Hermitian Hamiltonians, one generally does not obtain the variational bounds~\eqref{eq:variational_bounds} as can be observed in, \eg, the subspace-projected coupled-cluster method developed in Reference~\cite{Ekstrom:2019lss}.}
Furthermore, the General Ritz Theorem implies that the $\subspace{E}_n(\params)$ provide not only the variational bounds~\eqref{eq:variational_bounds} but also stationary approximations for these high-fidelity eigenvalues.
Adding another basis state to $X$ can only improve these approximations, which converge to the high-fidelity eigenvalues as the projected subspace approaches the high-fidelity space~\cite{Suzuki:1998bn}.

Although excited states can also be emulated, especially when adding excited-state snapshots to the trial wave function to improve the emulator's accuracy (see also Reference~\cite{Franzke:2021ofs}), we focus on ground-state properties and thus use only ground-state snapshots in the trial wave function.
For brevity, we will omit the subscripts henceforth.
To obtain the approximate ground-state wave function associated with $\subspace{E}(\params)$, one evaluates the Ritz vector $\ket{\psi(\params)} \approx \ritzbasis\coeffsopt(\params)$.
Expectations values of operators $O$ 
can then be straightforwardly computed using
\begin{align} \label{eq:ec_expectation_emulator}
    \braket{\psi(\params)| O(\params)|\psi(\params)} \approx \coeffsopt^\dagger(\params) \widetilde{O} \coeffsopt(\params),
\end{align}
with the subspace-projected $\widetilde{O}(\params)  = \ritzbasis^\dagger O(\params) \ritzbasis$. However, these expectation values generally do not provide variational bounds unless $O = H$ is the Hamiltonian, as discussed \new{(see, \eg, Figure~5 in Reference~\cite{Konig:2019adq} for emulated \isotope[4]{He} ground-state radii)}.

\subsection{Galerkin approach}

The reduced-order model~\eqref{eq:ritz_condition} can be alternatively derived via a Galerkin projection, as we will also see with the variational emulators for scattering in Section~\ref{sec:scattering-emulators}.
To this end, we construct the \emph{weak form} of the Schr{\"o}dinger Equation~\eqref{eq:generic_eigenvalue_problem} by left-multiplying it by an arbitrary test function $\bra{\testfunc}$ and asserting that
\begin{align} \label{eq:eigen_value_weak_full}
    \braket{\testfunc | H(\params) - E(\params) | \psi} = 0, \quad \forall \bra{\testfunc}.
\end{align}
%
If the weak form~\eqref{eq:eigen_value_weak_full} is satisfied for all $\bra{\testfunc}$ for a given set $\{E, \ket{\psi}\}$, then the set must also satisfy the Schrödinger Equation~\eqref{eq:generic_eigenvalue_problem}.
The proof of this statement can be obtained via a contrapositive: if the Equation~\eqref{eq:generic_eigenvalue_problem} were not satisfied, then one could find a $\bra{\testfunc}$ such that Equation~\eqref{eq:eigen_value_weak_full} is nonzero.

The weak form of the high-fidelity system is the starting point for deriving a reduced-order model.
Although Equation~\eqref{eq:eigen_value_weak_full} still operates in the large space in which $\ket{\psi}$ resides (cf. Figure~\ref{fig:illustration_rbm}), we can reduce its dimension by replacing $\ket{\psi} \to \ket{\trial\psi}$, where $\ket{\trial\psi}$ is defined in Equation~\eqref{eq:trial_general}.
With the degrees of freedom for $\ket{\psi}$ reduced, we enforce a less strict orthogonality condition: we select $\nbasis$ test functions $\testfunc_i$ and assert that the residual due to the trial wave function (cf. Figure~\ref{fig:illustration_rbm}) should be orthogonal to the subspace $\mathcal{Z}$ spanned by these test functions $Z = [\ket{\testfunc_1}, \dots, \ket{\testfunc_{\nbasis}}]$:
\begin{align}
    \left(H(\params)  - \subspace{E}(\params)  \right)\ket{\trial\psi} &\perp \mathcal{Z} \label{eq:eigenvalue_galerkin_geom},
\intertext{or likewise}
    \braket{\testfunc | H(\params)  - \subspace{E}(\params)  |\trial\psi} &= 0, ~~\forall \ket{\testfunc} \in \mathcal{Z}. \label{eq:eigenvalue_galerkin}
\end{align}
But replacing $\ket{\psi} \to \ket{\trial\psi}$ also implies that the true eigenvalue $E$ is in general not exactly reproduced unless 
$\mathcal{X}$ contains $\ket{\psi(\params)}$. 
Hence, we also had to apply the approximation $\subspace{E}\approx E$ in Equations~\eqref{eq:eigenvalue_galerkin_geom} and~\eqref{eq:eigenvalue_galerkin}.

In the Galerkin method, which is also known as the ``method of weighted residuals,'' the test and trial function bases are chosen to be equivalent;
\ie, $\mathcal{Z} = \mathcal{X}$.
The so-called Galerkin condition~\eqref{eq:eigenvalue_galerkin} is then equivalent to imposing that $\braket{\psi_i | H - \subspace{E} |\trial\psi} = 0$ holds for $i \in [1, \nbasis]$.
This yields a system of $\nbasis$ equations with $\nbasis$ unknowns $\coeffs$ and, together with the normalization condition, reduces to Equation~\eqref{eq:ritz_condition} obtained from the variational principle in Section~\ref{sec:eigen-emulators_variational}.
However, we stress that the test and trial function bases can be chosen differently (\ie, $\ket{\testfunc_i} \neq \ket{\psi_i}$), which makes the Galerkin method more general than the variational approach.
Note that the normalization condition does not affect the Galerkin condition~\eqref{eq:eigenvalue_galerkin} and can be implemented by normalizing the trial function.


\subsection{Emulator workflow and offline-online decomposition}

\begin{figure*}[tb]
    \centering
    \includegraphics[width=\textwidth]{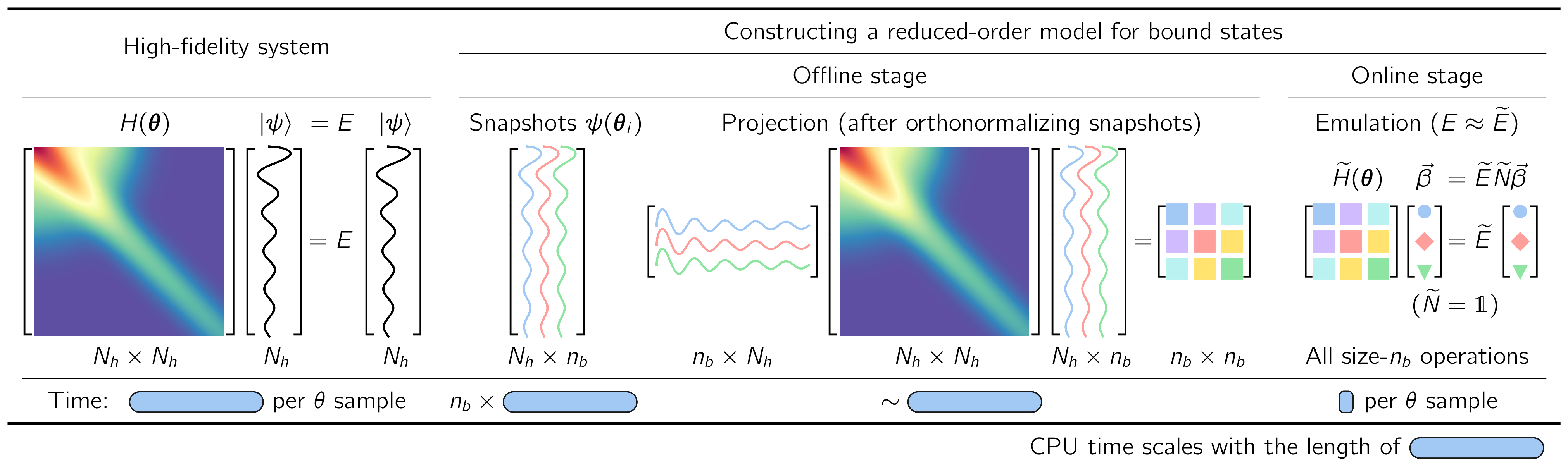}
    \caption{Illustration of the workflow for implementing fast \& accurate emulators, including a high-fidelity solver (left) and an intrusive, projection-based emulator with efficient offline-online decomposition (right), for sampling the (approximate) solutions of the Schrödinger Equation~\eqref{eq:generic_eigenvalue_problem} in the parameter space $\params$.
    For brevity, the figure assumes that the snapshots are orthonormalized during the offline stage such that $\subspace{N} = \identity$ in the emulator equation~\eqref{eq:ritz_condition}. 
    See the main text for details.
    }
    \label{fig:illustration_fom_rom}
\end{figure*}

Figure~\ref{fig:illustration_fom_rom} illustrates the workflow for implementing fast \& accurate emulators as described in Section~\ref{sec:eigen-emulators_variational}. 
The workflow involves
\begin{enumerate}
    \item a computational framework capable of reliably solving the high-fidelity system~\eqref{eq:generic_eigenvalue_problem},
    \item the snapshot-based trial wave function~\eqref{eq:trial_general} with the optimal coefficients (\ie, the weights) determined by the emulator Equation~\eqref{eq:ritz_condition}, and
    \item an efficient offline-online decomposition in which the computational heavy lifting is performed once \emph{before} the emulator is invoked.
\end{enumerate}

Several computational frameworks exist in nuclear physics (and quantum chemistry) for solving the few- and many-body Schrödinger Equation~\eqref{eq:generic_eigenvalue_problem}~\cite{Hergert:2020bxy}.
For illustration, Figure~\ref{fig:illustration_fom_rom} assumes that the high-fidelity solver performs a direct diagonalization of the $\nsimulator \times \nsimulator$ Hamiltonian in a chosen (truncated) model basis of length $\nsimulator$. 
The corresponding runtime $t_s$ per sampling point $\params_i$ is indicated by the width of the blue bar in Figure~\ref{fig:illustration_fom_rom}.
In nuclear physics, such approaches are referred to as Configuration Interaction (CI).
However, the following discussion will be independent of how the high-fidelity solutions of the Schrödinger Equation~\eqref{eq:generic_eigenvalue_problem} are obtained in practice.

Using the high-fidelity solver, one constructs a set of snapshots $\{\ket{\psi(\params_i)}\}_{i=1}^{\nbasis}$ in the truncated model basis to build the columns of the $\nsimulator \times \nbasis$ matrix $X$.
The runtime for this task is $\nbasis \times t_s$.
For simplicity, Figure~\ref{fig:illustration_fom_rom} assumes $\nbasis = 3$ and depicts the basis functions schematically in different colors.
This phase of the emulator needs to be completed only once before the emulator is invoked and is thus called the \emph{offline stage} as opposed to the \emph{online stage} of the emulator.
The predictions are made quickly and with little memory footprint in the online stage.

The appearance of the full-order Hamiltonian during the offline stage, where the projected Hamiltonian $\subspace{H}(\params) \equiv X^\dagger H(\params) X$ is computed (see Figure~\ref{fig:illustration_fom_rom}), implies that this class of projection-based emulators is \emph{intrusive} in nature. 
In general, intrusive emulators apply the basis expansions and projections to the operators implemented in the high-fidelity model~\cite{ghattas_willcox_2021}.
On the other hand, \emph{non-intrusive} emulators use only outputs of the
full-order solver without access to the full-order operators such as the Hamiltonian.
Non-intrusive emulators include Gaussian processes~\cite{rasmussen2006gaussian}, Dynamic Mode Decompositions~\cite{doi:10.1146/annurev-fluid-011212-140652,KutzDMDbook2016}, and other machine learning methods~\cite{raissi2019physics,CHEN2021110666,FRESCA2022114181}. 
More details on this classification scheme can be found in Section~8 in Ghattas \& Willcox~\cite{ghattas_willcox_2021}.

The emulator's efficiency greatly benefits from moving all size-$\nsimulator$ operations into the offline stage, which can easily be achieved for Hamiltonians $H(\params)$ with an affine parameter dependence. 
Those operators,
\begin{equation} \label{eq:H_affine}
    H(\params) = \sum_n h_n(\params) H_n,
\end{equation}
can be expressed as a sum of products of parameter-dependent functions $h_n(\params)$ and parameter-in\-de\-pendent operators $H_n$.
Note that the functions $h_n(\params)$ are only required to be smooth but not necessarily linear in $\params$.
The affine parameter dependence in Equation~\eqref{eq:H_affine} then allows one to store the subspace-projected operators $\subspace{H}_n = X^\dagger H_n X$ separately up front in the offline phase, from which
\begin{equation} \label{eq:Htilde_affine}
    \subspace{H}(\params) = \sum_n h_n(\params) \subspace{H}_n,
\end{equation}
can be efficiently constructed for each $\params_i$ during the \emph{online} stage 
to solve the emulator equation~\eqref{eq:ritz_condition}.
For instance, Hamiltonians derived from chiral EFT can be cast into the form~\eqref{eq:H_affine} due to their affine dependence on the low-energy couplings.
The runtime per sample $\params_i$ in the online phase is therefore typically just a small fraction of that of the high-fidelity solver, as depicted by the small blue box in Figure~\ref{fig:illustration_fom_rom}.
Likewise, emulating expectation values of other operators with an affine parameter dependence via Equation~\eqref{eq:ec_expectation_emulator} also benefits from this offline-online decomposition.
For non-affine operators, various hyperreduction methods have been developed to construct approximate affine representations~\cite{Quarteroni:218966,hesthaven2015certified}, including the empirical interpolation (EIM)~\cite{Barrault2004,Grepl2007,Chaturantabut2009,Chaturantabut2010} and gappy proper orthogonal
decomposition~\cite{gappyPOD2003,Carlberg2011gappyPOD}. 
See also References~\cite{Amsallem2010,AnCubature2008,Farhat2014,Benner20201} for hyperreduction methods that interpolate $X$ or $\subspace{H}(\params)$ directly, and References~\cite{GUO2018807,Zhang:2021jmi} for recent applications of machine learning tools for hyperreduction.

How should one choose the snapshots in the trial wave function~\eqref{eq:trial_general} effectively?
For relatively small parameter spaces, one can use Latin hypercube sampling to obtain space-filling snapshots or choose the snapshots in the proximity of the to-be-emulated parameter ranges, keeping $\nbasis \ll \nsimulator$ in practice.
A chosen set of snapshots expressed in the (truncated) model basis can be optimized by applying a singular value decomposition (SVD) or the closely related proper orthogonal decomposition (POD)~\cite{Gubisch2017} to the $\nsimulator \times \nbasis$ matrix $X$. 
One then creates a new set of snapshots from the (orthonormal) left-singular vectors associated with the singular values greater than a chosen threshold~\cite{hesthaven2015certified}.
This (optional) preprocessing step can be performed during the offline stage, as illustrated in Figure~\ref{fig:illustration_fom_rom}, thereby rendering the emulator equation~\eqref{eq:ritz_condition} an eigenvalue problem (\ie, $\subspace{N} = \identity$) and less sensitive to numerical noise.

The basis states of the trial wave function can also be obtained iteratively, using a greedy algorithm~\cite{Rozza2008,hesthaven2015certified,chen2017RBUQ}.
These algorithms estimate and then minimize the emulator's overall error by adding basis states (obtained from a high-fidelity solver) in the parameter space where the error is expected to be the largest.
Greedy algorithms require fast approximations of the emulator's error and terminate when either a requested error tolerance or a maximum number of iterations has been achieved.
Uncertainty quantification for reduced-order models has been studied in various contexts, including differential equations~\cite{chen2017RBUQ,hesthaven2015certified,Horger2017RBMeigenvalue} and nuclear physics problems~\cite{Sarkar:2021fpz,Bonilla:2022rph}.

\subsection{Illustrative example}
\label{sec:eigen-example}

The formal results so far in this Section can be illuminated by a simple example, which allows us to compare results from a snapshot-based emulator to more conventional approaches, such as direct diagonalization in a harmonic oscillator basis and Gaussian process emulation.
Let us define the system we would like to solve as a single particle with zero angular momentum in three dimensions and trapped in an anharmonic oscillator potential.
This example can be directly generalized to few- and many-body systems.
The potential operator is the sum of a conventional harmonic oscillator (HO) potential and a finite-range piece:
\begin{align}
    V(r; \params) = V_{\text{HO}}(r) + \sum_{n=1}^3 \param_n \exp(-r^2/\sigma_n^2), \label{eq:anharm_osc}
\end{align}
with $\sigma_n = [0.5, 2, 4] \; \mathrm{fm}$.
The potential~\eqref{eq:anharm_osc} has the affine structure defined in Equation~\eqref{eq:H_affine} for $\params$ and hence can be emulated rapidly after projecting into the snapshot basis during the offline stage.
Even the high-fidelity system considered here is still small enough to be solved quickly and accurately using a fine radial mesh on a standard laptop.
However, this provides an illuminating setting within which we can observe many qualities seen in more complicated scenarios.

\begin{figure}[tb]
\includegraphics{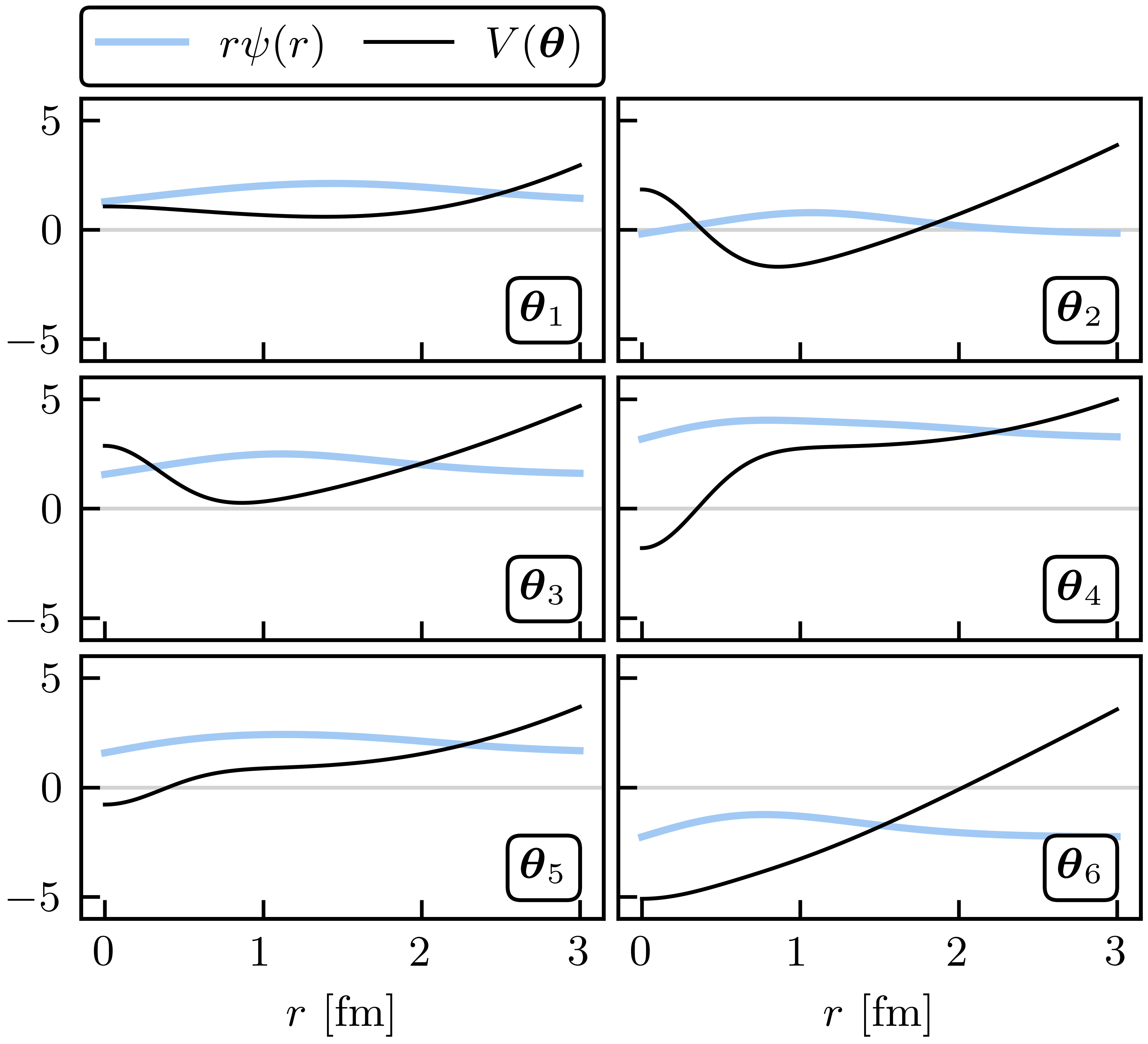}
\caption{
Basis functions for training the snapshot-based eigen-emulator. 
The black curves show the potential, and the blue curves show the wave functions as functions of the radial coordinate $r$.
The wave functions are offset vertically by their corresponding energies for clarity.
See the main text for details.
}
\label{fig:harm-osc}
\end{figure}

Following the MOR paradigm, we take snapshots of the high-fidelity wave function at various training parameters $\{\params_i\}$ and collect them into our basis $X$.
Here, we choose $\nbasis = 6$ training points randomly and uniformly distributed in the range $[-5, 5]$\,MeV for all $\theta_n$; 50 validation parameter sets are chosen within the same range.
The snapshots and the corresponding potentials are shown in Figure~\ref{fig:harm-osc}.
These snapshots are then used to construct the reduced-order system as in Equation~\eqref{eq:ritz_condition}.
All of this, and more, is made simple by the \texttt{EigenEmulator} Python class provided in the supplemental material~\cite{companionwebsite}.

Once the reduced system has been constructed and the affine structure of the Hamiltonian exploited to store the projected matrices during the offline stage, we can begin rapid emulation during the online stage.
To help provide a baseline to a common approach in nuclear physics, we provide an emulator constructed with the first $\nbasis=6$ HO wave functions as the trial basis $X$ in Equation~\eqref{eq:ritz_condition}.
We label this approach the HO emulator and the snapshot-based approach the reduced-basis method (RBM) emulator.
\new{See Reference~\cite{Melendez:2022kid} for a guide to the extensive literature on RBMs.}
One can emulate quantities with this HO approach via our \texttt{OscillatorEmulator} class~\cite{companionwebsite}.

\begin{figure}[tb]
\includegraphics{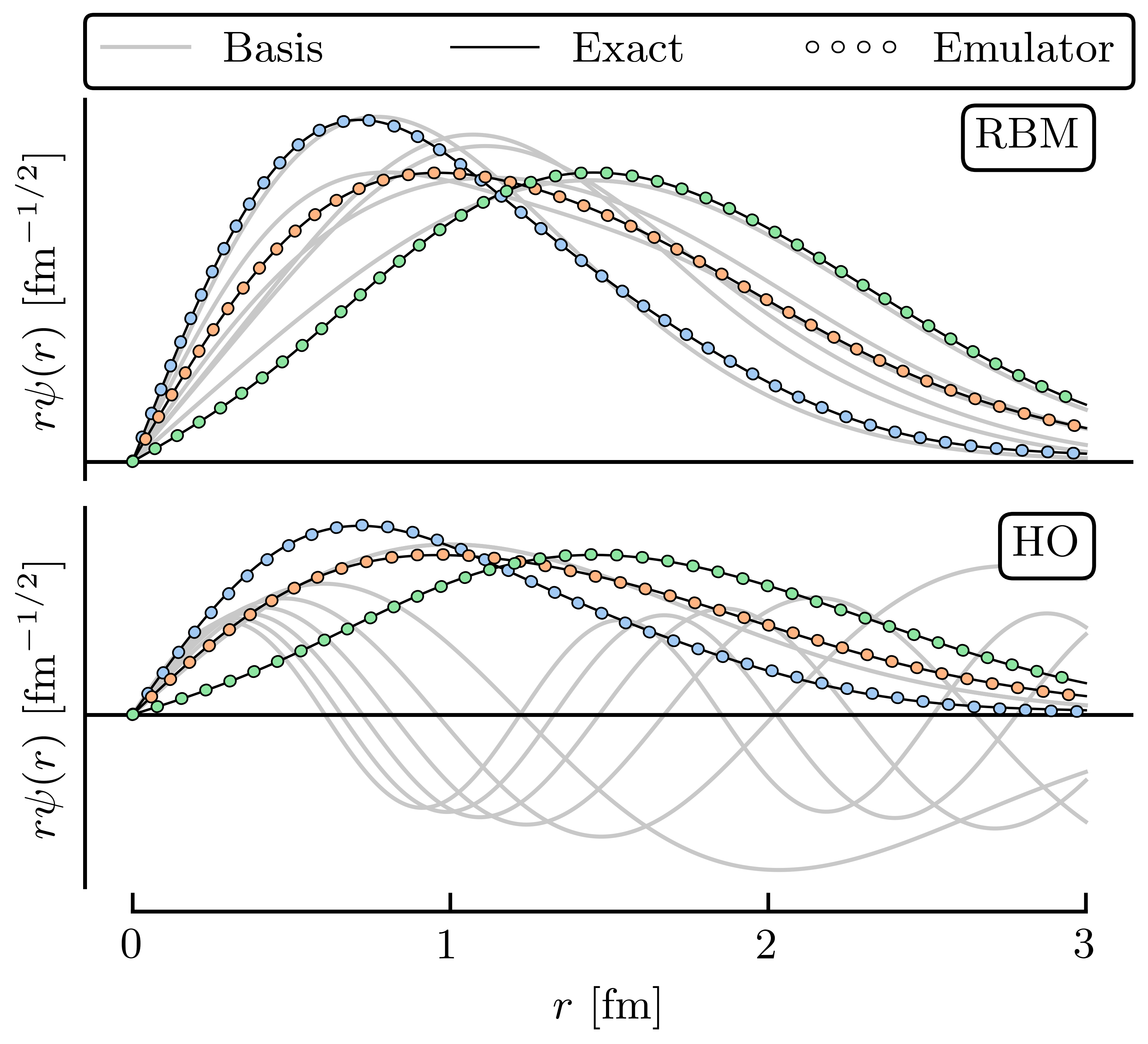}
\caption{
Emulated wave functions for the RBM emulator (top panel) and HO emulator (bottom panel) as a function of the radius.
The solid black lines represent the exact solution, and the dots represent the emulator result. 
The gray lines give the wave functions used to train the emulator.
See the main text for details.
}
\label{fig:eigen-emulator-wavefunctions}
\end{figure}

\begin{figure}[tb]
\includegraphics{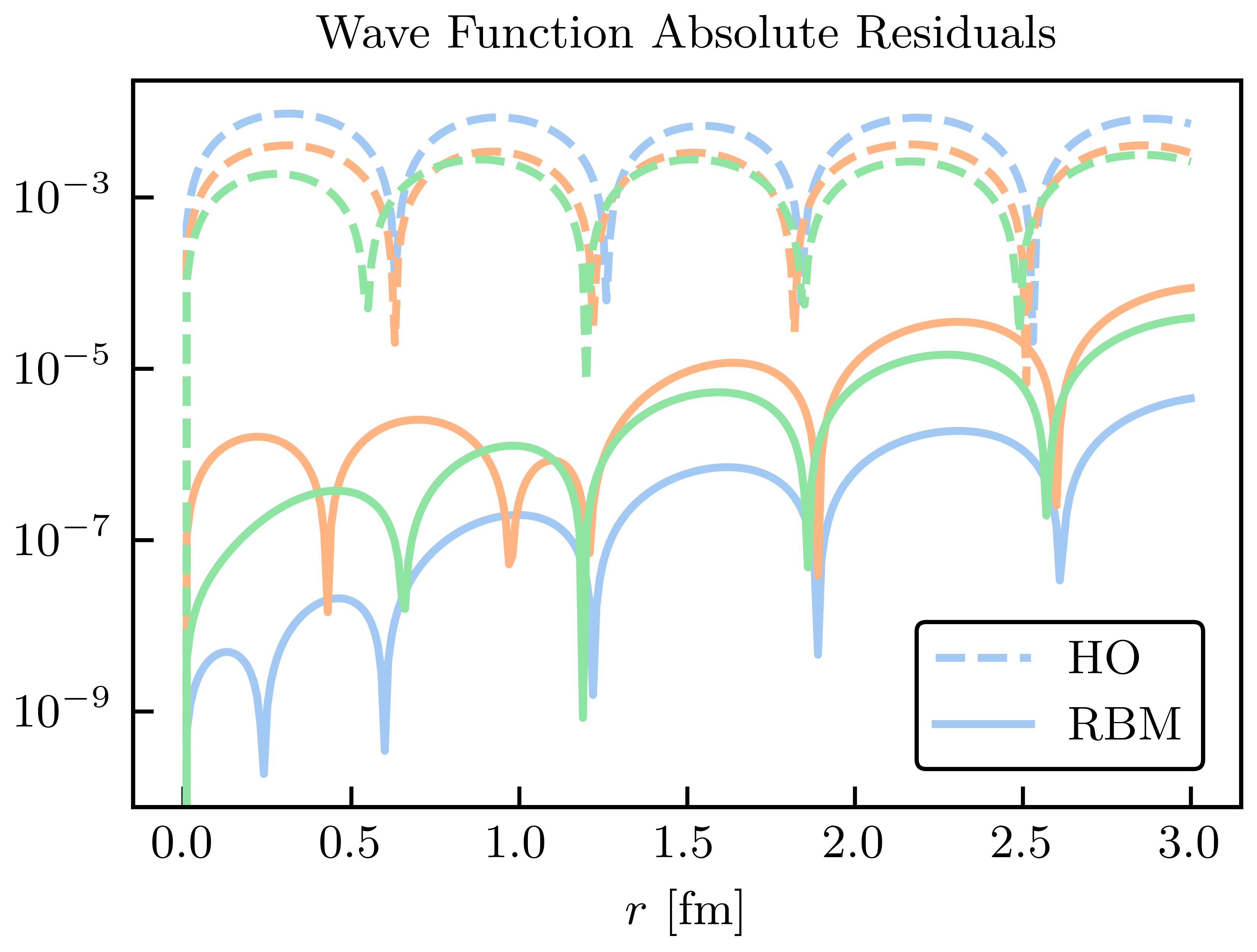}
\caption{
Absolute residuals of the emulated wave functions (in fm$^{-1/2}$) based on the RBM and HO emulator as a function of $r$. 
The emulator results are compared to the exact solutions. 
See the main text for details.
}
\label{fig:eigen-emulator-wavefunctions-residuals}
\end{figure}

For example, we take three of the validation parameter sets we sampled and compare the exact and emulated wave functions for both emulators.
Figure~\ref{fig:eigen-emulator-wavefunctions} shows the results.
The gray lines depict the $\nbasis$ wave functions used to create the reduced-order models, and the colored lines show the emulated results on top of the high-fidelity solutions (black lines).
Although both the reduced basis and HO basis are rich enough to capture the main effects of varying $\params$, the RBM emulator is much more effective at capturing the fine details of the wave function.
This can be seen in more detail in Figure~\ref{fig:eigen-emulator-wavefunctions-residuals}, where the absolute residuals of the RBM emulator are orders of magnitude smaller than those of the HO emulator. 
\new{The sensitivity of the emulator accuracy as $\nbasis$ is varied can be readily studied using the Python code provided on the companion website~\cite{companionwebsite}.}

The quality of the emulators can be understood by noting in Figure~\ref{fig:eigen-emulator-wavefunctions} that the basis functions of the RBM emulator match much more closely with the emulated wave functions than the HO emulator, whose wave functions have nodes not seen in the ground state (see the gray lines).
Thus, although the HO basis functions may be better at spanning the space of all possible wave functions, they are, in fact, a poor basis for spanning the set of all possible ground states as $\params$ are varied.
The RBM emulator constructs an extremely effective basis almost automatically, with minimal input required by the modeler.
This can prove particularly effective for cases where the system's complexity limits the quality of the basis that can be constructed from intuition or expertise alone.

Next, we discuss the emulation of bound-state observables.
Straightforward to emulate are the eigen-energies $E(\params)$, whose emulated values $\subspace{E}(\params)$ are the result of solving the emulator equation~\eqref{eq:ritz_condition}.
But as discussed in Section~\ref{sec:eigen-emulators_variational}, other observables associated with the operator $O$ can be emulated via $\braket{\psi(\params) | O | \psi(\params)} \approx \braket{\trial\psi(\params) | O | \trial\psi(\params)}$ using the $\trial\psi(\params)$ found from Equation~\eqref{eq:ritz_condition}.
We choose to show the results of emulating the radius-squared operator $R^2$, defined here to be
\begin{align}
    \braket{\psi | R^2 | \psi} = \frac{1}{\mathcal{N}} \int_0^\infty r^2\dd{r} r^2\psi^2(r) ,
\end{align}
\new{with the normalization $\mathcal{N} = \int_0^\infty r^2\dd{r} \psi^2(r)$.}
As stated previously, the bulk of the numerical effort in the evaluation of this matrix element is handled during the offline stage, where the integration is performed once,
\begin{align}
    \subspace{R}_{ij}^2 \equiv \braket{\psi_i | R^2 | \psi_j} = \frac{1}{\mathcal{N}} \int_0^\infty r^2 \dd{r} r^2\psi_i(r)\psi_j(r) ,
\end{align}
and then the online stage emulation can occur quickly via Equation~\eqref{eq:ec_expectation_emulator}; \ie, $\braket{\psi(\params) | R^2 | \psi(\params)} \approx \sum_{ij} \coeffsopt^{(i)}(\params)\subspace{R}_{ij}^2 \coeffsopt^{(j)}(\params)$.

For illustrative purposes, we continue our example using the trained RBM and HO emulators, but add a popular emulation tool to the discussion: Gaussian processes (GPs).
GPs are non-parametric, non-intrusive machine learning models for both regression \& classification tasks~\cite{rasmussen2006gaussian,Mackay:1998introduction,Mackay:2003information}.
Their popularity stems partly from their convenient analytical form and flexibility in effectively modeling various types of functions.
GPs benefit from treating the underlying set of codes as a black box~\cite{ghattas_willcox_2021}; as we will soon see, this is a double-edged sword.
We employ two independent GPs to emulate the ground-state energy and the corresponding radius expectation value.
Each GP uses a Gaussian covariance kernel and is fit to the observable values at the same values of $\params_i$ used to train the RBM emulator.
We use the maximum likelihood values for the hyperparameters.

\begin{figure}[tb]
\includegraphics[]{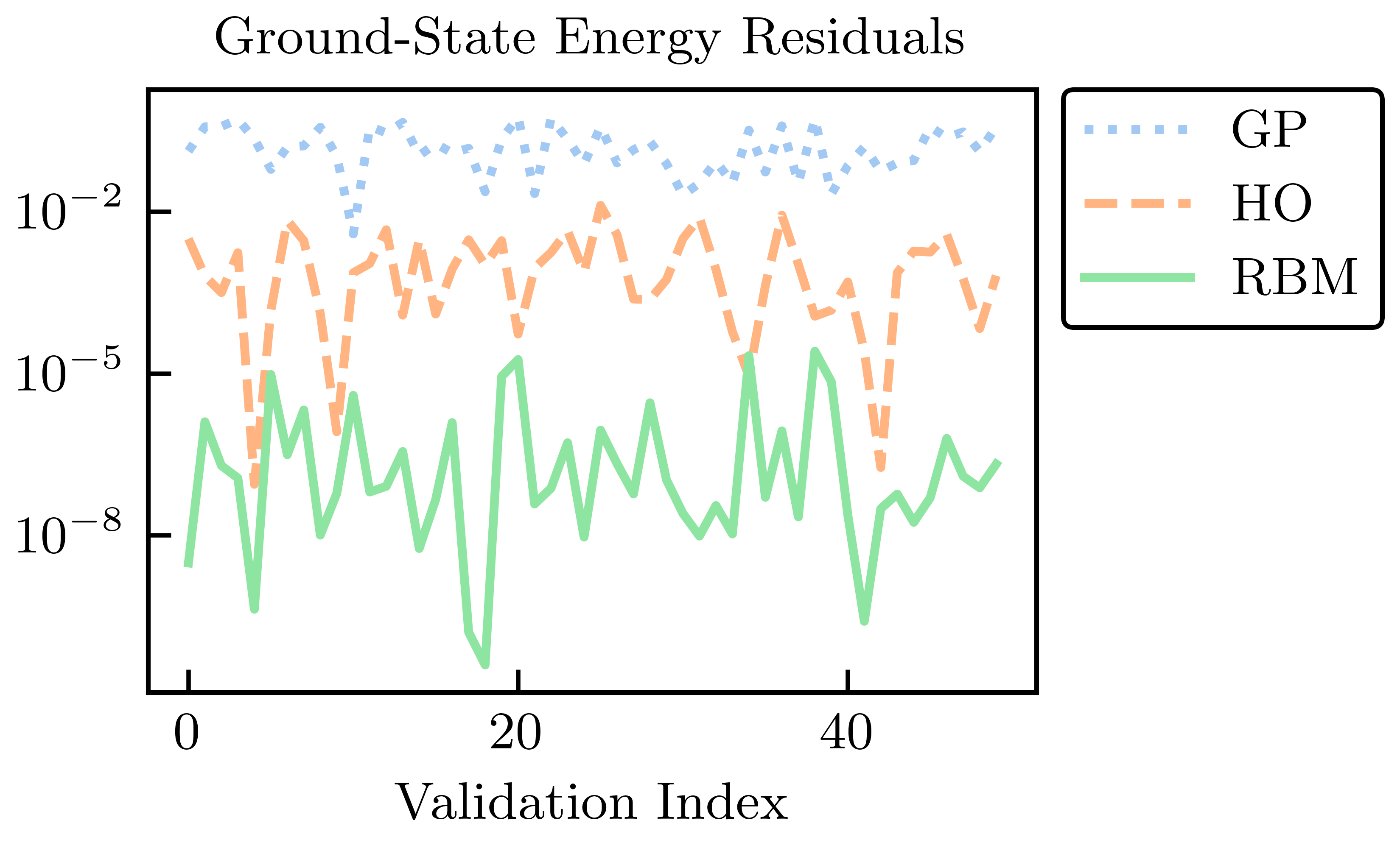}
\caption{
Absolute residuals in the energy (in MeV) at the 50 validation points for the RBM, HO, and GP emulators.
The validation points are chosen randomly from a uniform distribution within the same range as the training points.
See the main text for details.
}
\label{fig:energy-residuals}
\end{figure}

\begin{figure}[tb]
\includegraphics[]{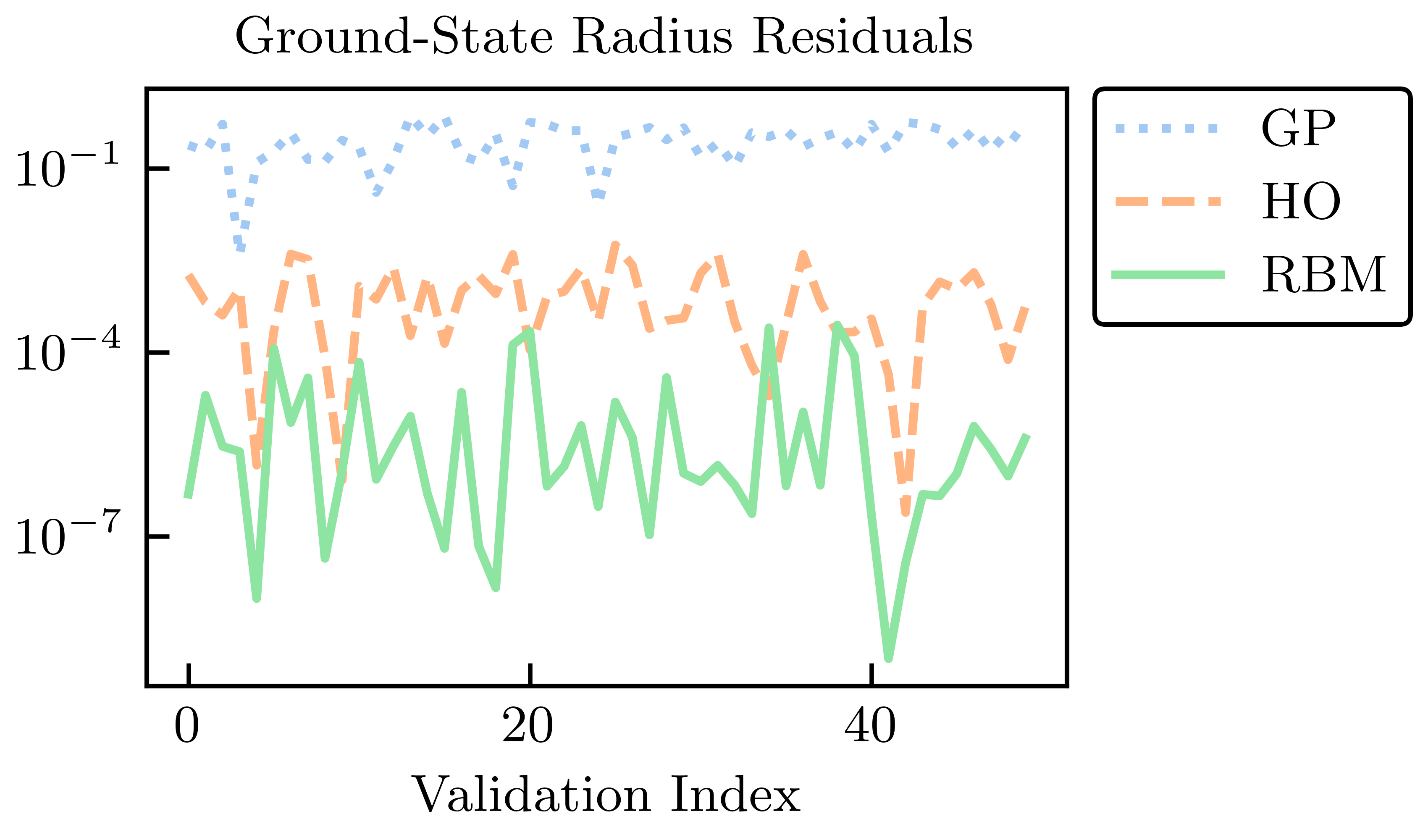}
\caption{
Similar to Figure~\ref{fig:energy-residuals} but for the root-mean-squared radius $\sqrt{\braket{R^2}}$ and in units of fm.
}
\label{fig:radius-residuals}
\end{figure}

The absolute residuals at the validation points for each of the RBM, HO, and GP emulators are shown in Figures~\ref{fig:energy-residuals} and~\ref{fig:radius-residuals} for the energy and radius, respectively.
Among these emulators, the GP emulators perform the worst, despite being trained on the values of the energies and radii themselves to perform this very emulation task.
Furthermore, its ability to extrapolate beyond the support of its training data is often poor unless great care is taken in the design of its kernel and mean function \new{(see Figures~1 and~2 in Reference~\cite{Konig:2019adq})}.
The GP suffers from what, in other contexts, could be considered its strength: because it treats the high-fidelity system as a black box (although some information can be conveyed via physics-informed priors for the hyperparameters), it cannot use the structure of the high-fidelity system to its advantage.
Note that the point here is not that it is impossible to find some GP that can be competitive with other RBM emulators after using expert judgment and careful (\ie, physics-informed) hyperparameter tuning.
Rather, we emphasize that with the reduced-order models, remarkably high accuracy is achieved \emph{without} the need for such expertise.

The HO emulator performs better than the GP emulator, but it was not ``trained'' \emph{per~se}, it was merely given a basis of the lowest six HO wave functions as a trial basis, from which a reduced-order model was derived.
However, the HO emulator can still outperform the GP emulator because it takes advantage of the \emph{structure} of the high-fidelity system: it is aware that the problem to be solved is an eigenvalue problem, for this is built into the emulator itself.
This feature permits a single HO emulator to emulate the wave function, energy, and radius simultaneously.

Coming in first in the comparison of the emulators' performances is the RBM emulator, which typically results in higher accuracies than the HO and GP emulators by multiple orders of magnitude.
The RBM emulator combines the best ideas from the other emulators.
Like the GP, the RBM emulator uses evaluations of the eigenvalue problem as training data.
However, its ``training data'' are \emph{curves} (\ie, the wave functions) rather than scalars (\eg, eigen-energies), like the GP is trained upon.
Like the HO emulator, the RBM emulator takes advantage of the structure of the system when projecting the high-fidelity system to create the reduced-order model.
With these strengths, the RBM emulator is highly effective in emulating bound-state systems, even with only a few snapshots \new{and far from the support of the snapshots (see Figures~1 and~2 in Reference~\cite{Konig:2019adq})}.
As we will see in Section~\ref{sec:scattering-emulators}, many of these strengths carry over to systems of differential equations.

\section{Model Reduction} \label{sec:model-reduction}

In this Section, we provide a more general discussion of variational principles and the Galerkin method as the foundations for constructing highly efficient emulators for nuclear physics \new{(see also Reference~\cite{Melendez:2022kid})}.
The general methods discussed here will be used as a springboard to develop emulators for the specific case of scattering systems in Section~\ref{sec:scattering-emulators}.

We consider (time-independent) differential equations that depend on the parameter vector $\params$ and
aim to find the solution $\trialfunc$ of
\begin{subequations} \label{eq:generic_differential_and_boundary}
\begin{align}
    D(\trialfunc; \params) & = 0 \quad \text{in } \Omega, \\
    B(\trialfunc; \params) & = 0 \quad \text{on } \Gamma,
\end{align}
\end{subequations}
where $\{D, B\}$ are differential operators and $\Omega$ is the domain with boundary $\Gamma$.
See Reference~\cite{Melendez:2022kid} for illustrative examples.
Here, we use the generic function $\trialfunc$ because different choices of $\trialfunc$ will be made in Section~\ref{sec:scattering-emulators}.
In what follows, we will discuss two related methods for constructing emulators from Equation~\eqref{eq:generic_differential_and_boundary}, 
which states the physics problem in a \emph{strong form} (\ie, Equation~\eqref{eq:generic_differential_and_boundary} holds for each point in the domain and on the boundary).
The first begins by finding a variational principle whose stationary solution implies Equation~\eqref{eq:generic_differential_and_boundary}.
The second instead constructs the corresponding \emph{weak form} of Equation~\eqref{eq:generic_differential_and_boundary}.

\subsection{Variational principles} \label{sec:variational}

Variational principles (VPs) have a long history in physics and play a central role in a wide range of applications; \eg, for deriving equations of motion using stationary-action principles and Euler--Lagrange equations in classical mechanics 
(see, \eg, Reference~\cite{Variational_Ritz_history_review_2012} for a historical overview).
Here, we use VPs as an alternate way of solving  the differential equations~\eqref{eq:generic_differential_and_boundary}. 

Variational principles are based on scalar functionals of the form 
\begin{equation} \label{eq:generic_functional}
    \action[\trialfunc] = \int_\Omega \dd{\Omega} F[\trialfunc] + \int_\Gamma \dd{\Gamma} G[\trialfunc],
\end{equation}
where $F$ and $G$ are differential operators. 
Many differential equations~\eqref{eq:generic_differential_and_boundary} can be solved by finding stationary solutions of a corresponding functional~\eqref{eq:generic_functional}; \ie, the solution $\trialfunc_\star$ that leads to $\delta\action[\trialfunc_\star] = 0$.

However, VPs can also lead straightforwardly to a reduced-order model.
This follows from the following trial ansatz
\begin{subequations} \label{eq:trial_general_subeq}
\begin{align}
    \ket{\trial\trialfunc} & \equiv \sum_{i=1}^{\nbasis} \coeff_i\ket{\trialfunc_i} = X\coeffs, \label{eq:trial_general_xi}\\
    X & \equiv
    \begin{bmatrix}
        \ket{\trialfunc_1} & \ket{\trialfunc_2} & \cdots & \ket{\trialfunc_{\nbasis}}
    \end{bmatrix},
\end{align}
\end{subequations}
with the to-be-determined coefficients vector $\coeffs$.
Rather than stipulate that $\delta\action = 0$ for any arbitrary variation $\delta\trialfunc$, we instead extract the optimal coefficients, $\coeffsopt$, as those for which $\action$ is stationary under variations in $\coeffs$: \footnote{For simplicity we consider $\trialfunc$ to be real a variable; for complex variables, independent variations $\delta\coeffs^\ast$ should be included in the discussion.}
\begin{equation} \label{eq:action_stationary_ansatz}
    \delta\action = \sum_{i=1}^{\nbasis} \frac{\partial\action}{\partial\coeff_i}\delta\coeff_i = 0.
\end{equation}
The general case would involve a numerical search for the solution to Equation~\eqref{eq:action_stationary_ansatz} but if $\action$ is quadratic in $\trialfunc$, as are all the examples considered here, then the solution can be determined exactly.
In this case, $\action$ can be written as
\begin{equation}
    \action[\coeffs] = \frac{1}{2}\coeffs^\trans A \coeffs + \vec{b}\cdot \coeffs + c
\end{equation}
for some matrix $A$, vector $\vec{b}$, and scalar $c$.
Symmetrizing the quadratic portion---if it is not already symmetric---by rewriting $A \leftarrow (A + A^\trans)/2$ can be desirable for numerical purposes.
It then follows that the optimal coefficients, $\coeffsopt$ are those for which
\begin{equation} \label{eq:coefficient_solve_quadratic}
    \delta\action = A_s\coeffs_\star + \vec{b} = 0,
\end{equation}
which can be solved for $\coeffsopt$ using standard linear algebra methods.
Solving for $\coeffsopt$ occurs only in a space of size $\nbasis$, the number of basis elements $\{\trialfunc_i\}_{i=1}^{\nbasis}$, rather than in the much larger space of $\trialfunc$ itself.
Therefore, as long as $\{\trialfunc_i\}_{i=1}^{\nbasis}$ approximately spans the space in which $\trialfunc$ lives, the trial function constructed by Equations~\eqref{eq:trial_general_xi} and~\eqref{eq:coefficient_solve_quadratic} will be both a fast \& accurate emulator of $\trialfunc$.

\new{Similar to the discussion in Section~\ref{sec:eigen-emulators_variational}, the matrix $A_s$ in Equation~\eqref{eq:coefficient_solve_quadratic} may be ill-conditioned and require regularization.
A nugget $\nu \ll 1$ can be added to the diagonal elements of $A_s$ to help stabilize the solution for $\coeffsopt$~\cite{Neumaier98solvingill-conditioned,engl1996regularization}.}

\subsection{Galerkin Emulators}

On the other hand, the Galerkin approach, also more broadly called the ``method of weighted residuals,'' relies on the \emph{weak} formulation of the differential equations~\eqref{eq:generic_differential_and_boundary}.
To obtain the weak form, the differential equation and boundary condition (in Equation~\eqref{eq:generic_differential_and_boundary}) are left-multiplied by arbitrary test functions $\testfunc$ and $\bar\testfunc$ and integrated over the domain and boundary, respectively, and then their sum is set to zero:
\begin{equation} \label{eq:weak_differential}
    \int_\Omega \dd{\Omega} \testfunc  D(\trialfunc) + \int_\Gamma \dd{\Gamma} \bar\testfunc  B(\trialfunc) = 0.
\end{equation}
If Equation~\eqref{eq:weak_differential} holds for all $\testfunc$ and $\bar\testfunc$, then Equation~\eqref{eq:generic_differential_and_boundary} must be satisfied as well.
The form of Equation~\eqref{eq:weak_differential} is often rewritten using integration by parts to reduce the order of derivatives and simplify the solution.
Importantly, the weak form has the integral form needed for our emulator application.
The weak form and its Galerkin projection are used extensively, \eg, in the finite element method; see References~\cite{zienkiewicz2013finite,Zienkiewicz2014finitesolid,Zienkiewicz2014finitefluid} for an in-depth study and examples.
For a discussion of the convergence properties of the Galerkin method, its relation to abstract variational problems, and other salient mathematical details, see References~\cite{hesthaven2015certified, Mikhlin_1967, Evans1996, Brenner:2008}.
Here, we follow the introduction of Galerkin methods as provided in Reference~\cite{zienkiewicz2013finite}.

Starting with the weak form, we can construct an emulator that avoids the need for an explicit variational principle.
It begins by first noting that substituting our trial function~Equation~\eqref{eq:trial_general} into $D(\trialfunc)$ and $B(\trialfunc)$ will not, in general, satisfy Equation~\eqref{eq:generic_differential_and_boundary} regardless of the choice of $\coeffs$.
Therefore, there will be some \emph{residual}, and the goal is to find the $\coeffsopt$ which minimizes that residual across a range of test functions $\testfunc$ and $\bar\testfunc$.
This system would be over-determined in the case of truly arbitrary test functions, so instead, we propose the test bases
\begin{align}
    \ket{\testfunc} & = \sum_{i=1}^{\nbasis} \delta\coeff_i\ket{\testfunc_i},
    \qquad
    \ket{\bar\testfunc} = \sum_{i=1}^{\nbasis} \delta\coeff_i\ket{\bar\testfunc_i},
\end{align}
where $\delta\coeff_i$ are arbitrary parameters, not related to the $\coeff_i$ in Equation~\eqref{eq:trial_general_xi}.
The $\delta\coeff_i$ will play the same role as those in Equation~\eqref{eq:action_stationary_ansatz}, namely as a bookkeeping method for determining the set of equations that are equivalently zero.
By enforcing that the residuals against these test functions vanish for arbitrary $\delta\coeff_i$, the bracketed expression in
\begin{align} 
    \delta\coeff_i \Bigl[\int_{\Omega} \dd{\Omega}  \testfunc_i  D(X\coeffsopt) +  \int_{\Gamma} \dd{\Gamma} \bar\testfunc_i  B(X\coeffsopt)
    \Bigl]= 0, \label{eq:weak_form_subspace} 
\end{align} 
is zero for all $i \in [1, \nbasis]$, from which the optimal $\coeffsopt$ are extracted.
Because this approximately satisfies the weak formulation, we have found an approximate solution to Equation~\eqref{eq:generic_differential_and_boundary}.

In a variety of cases \cite{zienkiewicz2013finite}, the subspace $\mathcal{Z}$ spanned by the test function basis is chosen to coincide with the subspace $\mathcal{X}$ spanned by the trial function basis $X$; \ie, $\mathcal{Z} = \mathcal{X}$.
This particular choice is known as \emph{the} Galerkin method, but it is sometimes further specified as the Ritz--Galerkin or Bubnov--Galerkin methods.
The Galerkin method is more general than the variational methods described in Sec.~\ref{sec:variational} because the test space need not be equivalent to the trial space (\ie, $\mathcal{Z} \neq \mathcal{X}$).
In these cases, the approach is described as the Petrov--Galerkin method~\cite{zienkiewicz2013finite};
this can result in more efficient emulators for some differential equations~\cite{Zienkiewicz2014finitefluid}.

\section{Scattering Emulators} \label{sec:scattering-emulators}

In this Section, we describe various reduced-basis emulators one could construct for quantum scattering systems.
Throughout, we note how the variational principles used to construct emulators in recent works are related~\cite{Furnstahl:2020abp,Drischler:2021qoy,Zhang:2021jmi, Melendez:2021lyq,KVP_vs_NVP:2022}.
We also describe how each of the results from VPs could instead be derived from Galerkin projections.

For scattering problems, the Schrödinger equation~\eqref{eq:generic_eigenvalue_problem} is 
no longer an eigenvalue problem. 
The task is to solve the differential equation for the wave function at a given energy $E$ rather than searching for discrete energies with normalizable wave functions.
Differential equations are well studied in the field of MOR, where parametric reduced-order models have been constructed with great success across a multitude of fields~\cite{Benner2020Volume1DataDriven,Benner2020Volume3Applications}.
This is a relatively mature field whose formal results are quite extensive.
For example, UQ for the RBM has been well studied, along with the development of effective algorithms for choosing the best training points~\cite{chen2017RBUQ,hesthaven2015certified,HUYNH2007473,Rozza2008}.

One can formulate the Schrödinger equation in multiple ways, including any flavor of Lippmann-Schwinger (LS) integral equation (which builds in boundary conditions) or as a differential equation in either homogeneous or inhomogeneous form.
This freedom, along with the freedom of trial and test bases for the Galerkin projection, leads to multiple alternative emulators that one could construct for quantum scattering systems.
\new{For simplicity, we restrict our discussion to two-body scattering for Hermitian Hamiltonians. (See, however, Section~\ref{sec:higher-body-continuum} for an extension to higher-body systems.)}
As a concise reference, we provide Table~\ref{tab:scattering-vps-galerkin} to show the connections between the fundamental differential or integral equations, variational principles, and Galerkin projections.
This section thus provides multiple distinct examples of using Galerkin projections to create emulators, which may prove useful to newcomers wishing to apply model reduction to their own systems, and ends with an example for an emulator applied to a separable potential.
%

\begin{table*}[tb]
\renewcommand{\arraystretch}{1.7}
\caption{Description of common variational principles (VPs) in quantum scattering, and how to relate them to a Galerkin projection.
The quantities are defined as the free wave function $\ket{\phi}$, the full wave function $\ket{\psi}$, the scattered wave function $\ket{\chi}$, and the reactance matrix $K$ along with its on-shell form $K_E$.
Tildes denote trial quantities.
The expressions for the Newton VP are written in operator rather than scalar form; \new{any matrix element can be made individually stationary (see Section~\ref{sec:newton} for details)}.
To compute the weak form of the Schwinger and Newton VPs, one must first left multiply by $V(\params)$ and $G_0$, respectively, before orthogonalizing against the test basis.
The rightmost column specifies whether a constraint for the trial wave function has to be imposed (\eg, using a Lagrange multiplier $\lambda$).
}
\label{tab:scattering-vps-galerkin}
\centering
\renewcommand{\arraystretch}{2}%
\begin{tabular}{p{1.15cm}ccccc}
\toprule
\multicolumn{2}{c}{Variational Principle} & \multicolumn{4}{c}{Galerkin Projection Information}\\
\cmidrule(lr){1-2}\cmidrule(lr){3-6}
Name & Functional for $K$ & Strong Form & Trial Basis & Test Basis & \new{Constrained?} 
\\
\cmidrule(lr){1-6}
Kohn ($\lambda$) & $\trial K_E + \braket{\trial\psi | H - E | \trial\psi}$ & $H\ket{\psi} = E\ket{\psi}$ & $\ket{\psi_i}$ & $\bra{\psi_i}$ & Yes
\\[10pt]

Kohn (No $\lambda$)
&
{
$
\begin{aligned}
\braket{\trial\chi | H - E | \trial\chi} +
\braket{\phi | V | \trial\chi} \\
+\braket{\phi | H - E | \phi} +
\braket{\trial\chi | V | \phi}
\end{aligned}
$
}
& $[E-H]\ket{\chi} = V\ket{\phi}$ & $\ket{\chi_i}$ & $\bra{\chi_i}$ & No
\\[15pt]

Schwinger &
{
$
\begin{aligned}
\braket{\trial\psi | V | \phi} + \braket{\phi | V | \trial\psi} \\
- \braket{\trial\psi | V - V G_0 V | \trial\psi}
\end{aligned}
$
}
& $\ket{\psi} = \ket{\phi} + G_0 V\ket{\psi}$ & $\ket{\psi_i}$ & $\bra{\psi_i}$ & No
\\[18pt]

Newton &
{$
\begin{aligned}
V + V G_0 \trial K + \trial K G_0 V \\
 - \trial K G_0 \trial K + \trial K G_0 V G_0 \trial K
\end{aligned}$
}
& $K = V + VG_0 K$ & $K_i$ & $K_i$ & No \\
\bottomrule
\end{tabular}
\end{table*}

\subsection{Constrained Kohn Emulators} \label{sec:kohn-lagrange}

The Kohn variational principle (KVP)~\cite{Kohn:1948col,Kohn:1951zz} is one of the most well-known VPs for quantum scattering systems.
Here we focus on the KVP flavor that relates a trial wave function to the reactance matrix $K$.
However, alternative flavors exist for other matrices such at $T^{\pm}$ and $S$ (see Section~\ref{sec:kohn-general}).
Analogously to the Ritz VP for bound states, the KVP then allows us to guess effective wave functions by finding those that make the KVP stationary.
This Section will discuss a style of KVP emulator that relies on the homogeneous Schrödinger equation, which requires a normalization constraint during emulation; an alternative style without such a constraint will be discussed in Section~\ref{sec:kohn-no-lagrange}.

We start with a rescaled version of the KVP discussed in Reference~\cite{Furnstahl:2020abp}:
\begin{align} \label{eq:kohn-psi}
    \mathcal{K}[\trial\psi] = \trial K_E + \braket{\trial\psi | H - E | \trial\psi},
\end{align}
where $\ket{\trial\psi}$ is the trial wave function (denoted by $\ket{\trial\trialfunc}$ in Section~\ref{sec:model-reduction}) and $\trial K_E \equiv \sum_i \coeff_i K_{E,i}$ the associated on-shell trial $K$ matrix with on-shell energy $E = q^2/2\mu$.
This flavor of KVP applies when $\psi$ satisfies the asymptotic normalization condition in position space
\begin{align} \label{eq:psi-normalization}
    \psi_\ell(r) \xrightarrow[r\to\infty]{} j_{\ell}(qr) + n_{\ell}(qr) \tan \delta_{\ell},
\end{align}
%
where $\phi(r) = j_\ell(qr)$ is the (regular) free-space wave function, and $j_\ell(qr)$ and $n_\ell(qr)$ are spherical Bessel and Neumann functions, respectively.%
\footnote{\new{
We focus here on examples with real-valued potentials and without long-range Coulomb interactions. 
Cases with complex-valued potentials and/or the Coulomb interaction may be analyzed in similar ways; relevant discussions specific to Kohn emulators can be found in References~\cite{Furnstahl:2020abp,Drischler:2021qoy}.}}
Note that we define the on-shell $K_E$ matrix as 
\begin{equation}
    K_E = -\frac{\tan\delta_\ell}{2\mu q},
\end{equation}
which differs from the convention in Reference~\cite{Furnstahl:2020abp}.
The KVP is stationary about exact solutions $\psi$, such that $\mathcal{K}[\psi + \delta\psi] = K_E + \mathcal{O}(\delta K)^2$.

Equation~\eqref{eq:kohn-psi} can be cast into the form of the generic functional~\eqref{eq:generic_functional} by noting that, in position space,
\begin{align} \label{eq:kohn-psi-surface-functional}
    \int_\Gamma \dd{\Gamma} G[\trial\psi] \to \!\left. G[\trial\psi] \right|_{r=0}^\infty & \equiv \!\left. \frac{W(r\phi, r\trial\psi; r)}{2\mu} \right|_{r=0}^\infty \notag\\
    & = \trial K_E,
\end{align}
which has defined the surface functional $G$ in Equation~\eqref{eq:generic_functional} and where we have used the Wronskian
\begin{align} \label{eq:wronskian}
    W(\phi, \psi; r) \equiv \phi(r) \psi'(r) - \phi'(r)\psi(r).
\end{align}
Both $r\phi(r)$ and $r\trial\psi(r)$ vanish at $r=0$ so only the limit of $r\to\infty$ contributes, from which we can use Equation~\eqref{eq:psi-normalization} when evaluating Equation~\eqref{eq:kohn-psi-surface-functional}.

Because the Schrödinger equation is a linear, homogeneous differential equation, the normalization of $r\psi(r)$ is proportional to its derivative at, say, $r=0$.
Therefore, a constraint on the normalization of $\psi$ is equivalent to a boundary condition on $\psi'$.
However, to satisfy this boundary condition we must include a constraint on Equation~\eqref{eq:kohn-psi} if we are to ensure that the trial function $\trial\psi$ continues to satisfy the normalization condition of Equation~\eqref{eq:psi-normalization}.
If we assume that each snapshot $\psi_i$ satisfies Equation~\eqref{eq:psi-normalization}, then
\begin{align}
    \trial\psi_\ell(r) = \left[\sum_i \coeff_i\right] j_{\ell}(qr) + n_{\ell}(qr) \sum_i \coeff_i \tan \delta_{\ell,i},
\end{align}
whose first term implies that we must impose the constraint $\sum_i \coeff_i = 1$.

We are now in a position to find the $\coeffs$ that make the Equation~\eqref{eq:kohn-psi} stationary.
If we insert the definition of $\trial\psi$ and $\trial K$ into Equation~\eqref{eq:kohn-psi}, along with the Lagrange multiplier, we have (with repeated indices indicating summations)
\begin{align} \label{eq:kohn-psi-reduced}
    \mathcal{K}[\coeffs] = \coeff_i K_{E,i} + \frac{1}{2}\coeff_i \dU_{ij} \coeff_j + \lambda \!\left[\sum_i \coeff_i - 1\right],
\end{align}
where we define $V_i = V(\params_i)$ and
\begin{align} \label{eq:delta_u_tilde}
    \dU_{ij} & \equiv \braket{\psi_i | H - E | \psi_j} + (i\leftrightarrow j)\notag\\
    & = \braket{\psi_i | V(\params) - V_j | \psi_j}  + (i\leftrightarrow j) .
\end{align}
In the second line we have used that the $\ket{\psi_j}$ are eigenstates with the corresponding $V_j$.
If $V(\params)$ is affine in $\params$, then $\dU$ can be projected once in the emulator's offline stage, and reconstructed quickly during the online stage.

Now we can follow the steps outlined in Section~\ref{sec:variational} to determine $\coeffsopt$.
Taking the gradient of Equation~\eqref{eq:kohn-psi-reduced} with respect to $\coeffs$ and setting it equal to 0 yields
\begin{align} \label{eq:kohn-psi-gradient}
    \vec{K}_{E} + \dU \coeffsopt + \lagmult = 0,
\end{align}
where $\vec{K}_E$ are the $\nbasis$ on-shell $K$-matrices used to train the emulator, and $\coeffsopt$ are the optimal coefficients of the trial wave function.
The gradient with respect to $\lambda$ simply returns the constraint.
This system can be solved via the system of equations
\begin{align} \label{eq:kohn-psi-coeff-solution}
    \begin{bmatrix}
        \dU & \vec{1}\,{} \\
        \vec{1} \, {}^\intercal & 0\,{}
    \end{bmatrix}
    \begin{bmatrix}
        \coeffsopt \, {} \\
        \lagmult \, {}
    \end{bmatrix}
    =
    \begin{bmatrix}
        -\vec{K}_E \\
        1
    \end{bmatrix},
\end{align}
where $\vec{1}$ is an $\nbasis \times 1$ vector of ones.
If the $\nbasis$ number of basis functions is much smaller than the size of $\psi$, then Equation~\eqref{eq:kohn-psi-coeff-solution} can be a highly computationally efficient emulator for scattering systems and requires little computer memory to store.
The on-shell $K$ matrix can then be emulated via
\begin{align} \label{eq:kohn-psi-K-emulator}
    K_E \approx \mathcal{K}[\coeffsopt] = \coeffsopt \cdot \vec{K}_E + \frac{1}{2}\coeffsopt^\trans \dU \coeffsopt ,
\end{align}
whose operations all occur quickly in the size-$\nbasis$ space during the online stage.

The derivation above followed closely that of References~\cite{Furnstahl:2020abp,Drischler:2021kxf}, but one could instead arrive at exactly Equation~\eqref{eq:kohn-psi-coeff-solution} from a Galerkin projection~\cite{Bonilla:2022rph}.
Rather than begin with the VP, we start here with (the \emph{strong form} of) the homogeneous Schrödinger equation, \ie,
%
    $H(\params) \ket{\psi} = E \ket{\psi}$.
%
To construct the weak form, we multiply with a test function $\ket{\testfunc}$ and assert that the residual vanishes:
\begin{align} \label{eq:kohn-weak-initial}
    \braket{\testfunc | H - E | \psi} = 0, \qquad \forall \ket{\testfunc}.
\end{align}
To make explicit the boundary conditions, we make use of the relation 
\begin{multline} \label{eq:int-by-parts-wronskian}
    0=\braket{\testfunc | H - E | \psi} \\= \braket{\testfunc | \overleftarrow{H} - E | \psi} - \left.\frac{W(r\testfunc, r\psi; r)}{2\mu} \right|_{r=0}^{\infty},
\end{multline}
where we have again used the Wronskian from Equation~\eqref{eq:wronskian}, and assigned $\overleftarrow{H}$ as the operator acting, after integration by parts, on $\langle \testfunc | $ instead of  $|\psi\rangle$.  
By adding Equations~\eqref{eq:int-by-parts-wronskian} and~\eqref{eq:kohn-weak-initial}, we have
\begin{multline} \label{eq:kohn-weak-full}
    \braket{\testfunc | H - E | \psi} + \braket{\testfunc | \overleftarrow{H} - E | \psi} \\ = \left.\frac{W(r\testfunc, r\psi; r)}{2\mu} \right|_{r=0}^{\infty}.
\end{multline}
%
This is the weak form of the homogeneous Schrödinger equation that we will use to construct the emulator, although the asymptotic normalization condition~\eqref{eq:psi-normalization} still needs to be enforced.
This will be imposed via a Lagrange multiplier after inserting our trial basis.

Now that we have a weak form, the next step to construct the reduced-order model equations is to define our trial and test bases to project the weak form into the finite space of these bases.
To align with the Kohn emulator from the variational argument above, we choose the trial and test basis to be identical as snapshots $\psi_i$.
Then we can evaluate
\begin{align}
    \left.\frac{W(r\psi_i, r\psi_j; r)}{2\mu} \right|_{r=0}^{\infty} = K_j - K_i
\end{align}
and thus, it follows after including the Lagrange multiplier that
\begin{align} \label{eq:kohn-weak-reduced-unfinished}
    \lambda + \Big[\braket{\psi_i | H - E | \psi_j} & + \braket{\psi_i | \overleftarrow{H} - E | \psi_j}\Big] \coeff_j \notag\\
    & \hspace{-0.4in} = \sum_j \coeff_j K_j - K_i \sum_j \coeff_j.
\end{align}
The sum in the rightmost term can be evaluated using the constraint $\sum_j \coeff_j=1$, and we can make the redefinition $\lambda' \equiv \lambda - \sum_j \coeff_j K_j$ without impacting the solution because this term does not depend on $i$.
Thus, we have
\begin{align} \label{eq:kohn-weak-reduced}
    \lambda' + \vec{K}_E + \dU \coeffsopt = 0,
\end{align}
which is exactly Equation~\eqref{eq:kohn-psi-gradient} found by making the KVP stationary.
This simplification can be understood by noting that if $\{\coeffsopt, \lambda_\star\}$ satisfy Equation~\eqref{eq:kohn-weak-reduced-unfinished}, then we know that $\{\coeffsopt, \lambda_\star'\}$ is the unique solution to Equation~\eqref{eq:kohn-weak-reduced}.
Therefore, we can solve Equation~\eqref{eq:kohn-weak-reduced} to obtain $\coeffsopt$ rather than Equation~\eqref{eq:kohn-weak-reduced-unfinished}.
\new{In conclusion, using the Galerkin projection of the homogeneous Schrödinger equation with trial and test bases of $\psi_i$, we were able to obtain the same coefficients as the KVP in Equation~\eqref{eq:kohn-psi-coeff-solution}, which yield the same on-shell $K$ matrix when used in Equation~\eqref{eq:kohn-psi-K-emulator}.}

\subsection{Unconstrained Kohn Emulators}
\label{sec:kohn-no-lagrange}

The Kohn emulators from Section~\ref{sec:kohn-lagrange} start with the homogeneous Schrödinger equation, which does not enforce any specific normalization of the wave function; hence this requirement needs to be enforced at the time of emulation.
This effectively takes the $\nbasis$ degrees of freedom $\{\psi_i\}$---which were potentially costly to obtain---and reduces the degrees of freedom to $\nbasis-1$.
\new{However, one can instead build in the normalization from the very start, thus removing the need to constrain our basis via $\sum_{j=1}^{\nbasis} \coeff_j = 1$ during emulation.
The unconstrained emulator is fundamentally different from any approach that constrains the coefficients (\eg, explicit substitution of $\coeff_1 = 1 - \sum_{j=2}^{\nbasis} \coeff_j$), regardless of if a Lagrange multiplier is explicitly used as in Section~\ref{sec:kohn-lagrange}.}
This is the topic of the current section.

The full wave function $\ket{\psi}$ can be written as the sum of the free wave function $\ket{\phi}$ and the scattered wave $\ket{\chi}$, that is, $\ket{\psi} = \ket{\phi} + \ket{\chi}$.
Thus, we can rewrite the KVP as
\begin{align}
    \mathcal{K} & = K + \braket{\psi | [H - E] | \psi} \notag\\
    & = K + \braket{\chi | [H - E] | \chi} + \braket{\phi | [H - E] | \chi} 
      \notag \\
    & \qquad~ + \braket{\phi | [H - E] | \phi} + \braket{\chi | [H - E] | \phi} \notag\\
    & = \braket{\chi | [H - E] | \chi} + \braket{\phi | V | \chi} \notag\\
    & \quad\, + \braket{\phi | [H - E] | \phi} + \braket{\chi | V | \phi} , \label{eq:kohn-no-lagrange}
\end{align}
where we used (via integration by parts)
\begin{equation}
    \braket{\phi | [H - E] | \chi} - \braket{\phi | [\overleftarrow{H} - E] | \chi} = -K.
\end{equation}
We choose our trial function as 
$\ket{\trial\chi}$, which always enforces the normalization condition $\ket{\trial\psi} = \ket{\phi} + \ket{\trial\chi}$, and so
no additional constraint needs to be included in the variational principle.

Now we can construct the set of linear equations that makes Equation~\eqref{eq:kohn-no-lagrange} stationary in $\ket{\trial\chi} = \sum_{i} \coeff_i \ket{\chi_i}$.
By taking the gradient with respect to $\coeff_i$, we find \new{
\begin{align} \label{eq:kohn-no-lagrange-solution}
    \Omega \coeffsopt = \vec{\omega},
\end{align}
where
\begin{subequations}
\begin{align} \label{eq:kohn-unconstrained-reduced-matrices}
    \Omega_{ij} & = \braket{\chi_i | [E - H] | \chi_j}, \\
    \omega_i & = \braket{\chi_i | V | \phi},
\end{align}
\end{subequations}
%
which is the set of equations used to obtain $\coeffsopt$.
The matrix elements $\Omega_{ij}$ can be evaluated with the help of
}
\begin{align}
    [E - H] \ket{\chi_j} & = [E - H_j] \ket{\chi_j} + [V_j - V] \ket{\chi_j} \notag\\
    & = V_j \ket{\phi} + [V_j - V]\ket{\chi_j},
\end{align}
with $H_j = H(\params_j)$ and $V_j = V(\params_j)$.

An equivalent approach follows from a Galerkin orthogonalization procedure.
We begin by writing the homogeneous Schrödinger equation in inhomogeneous form using $\ket{\psi} = \ket{\phi} + \ket{\chi}$:
\begin{align}
    [E - H] \ket{\chi} = V \ket{\phi}.
\end{align}
We can construct the weak form by multiplying by a generic test function $\ket{\testfunc}$, which yields
\begin{align}
   \braket{\testfunc | E - H | \chi} = \braket{\testfunc | V | \phi}.
\end{align}
Next, we insert the trial function $\ket{\trial\chi}$ and choose the test basis of $\{\ket{\chi_i}\}_i$, which is the same as the trial basis.
This yields a reduced weak form that is identical to Equation~\eqref{eq:kohn-no-lagrange-solution}.

We have shown that the coefficients $\coeffsopt$ found via the appropriate Galerkin procedure aligns exactly with the KVP\@.
However, we can go one step further and in fact derive an estimate for the $K$ matrix that is equivalent to $\mathcal{K}[\coeffsopt]$.
By inserting the optimal coefficients into $K\ket{\phi} = V\ket{\psi}$,
\begin{align}
    \braket{\phi' | K | \phi} & \approx \braket{\phi' | V | \phi} + \braket{\phi' | V | \trial\chi} \notag \\
    & =  \braket{\phi' | V | \phi} 
    \notag \\
    & \quad\null + \sum_{ij} \braket{\phi' | V | \chi_i} \bigg[ (\Omega^{-1})_{ij} \braket{\chi_j | V | \phi}\bigg],
\end{align}
with the factors in brackets equating to $\coeff_i$ using Equation~\eqref{eq:kohn-no-lagrange-solution}.
The equivalence to the KVP is demonstrated in References~\cite{Takatsuka1981SchwingerKohnRelationship,Takatsuka1981Scattering}.

\subsection{Schwinger Emulators} \label{sec:schwinger}

The Schwinger variational principle (SVP) is given by~\cite{Takatsuka1981SchwingerKohnRelationship}
\begin{equation} \label{eq:schwinger-vp}
    \mathcal{K}[\trial\psi] = \braket{\trial\psi | V | \phi} + \braket{\phi | V | \trial\psi} - \braket{\trial\psi | V - V G_0 V | \trial\psi},
\end{equation}
\new{where $G_0$ is the Green's operator.}
This too has the stationary property $\mathcal{K}[\psi + \delta\psi] = K + \mathcal{O}(\delta K)^2$ when $\psi$ is a wave function satisfying the LS equation.
Following the MOR philosophy and inserting a trial function $\trial\psi$, the stationary condition becomes
\begin{equation} \label{eq:schwinger-linear}
    W\coeffsopt = \vec{w},
\end{equation}
where
\begin{subequations}
\begin{align}
W_{ij} & = \braket{\psi_i | V - V G_0 V | \psi_j} \\
w_i & = \braket{\psi_i | V | \phi},
\end{align}
\end{subequations}
for all $i \in[ 1, \dots, \nbasis]$.

The system of equations~\eqref{eq:schwinger-linear} can also be determined by a Galerkin projection procedure.
In this case, we start with the LS equation for wave functions,
\begin{equation} \label{eq:ls-wave-function}
    \ket{\psi} = \ket{\phi} + G_0 V \ket{\psi},
\end{equation}
and create a weak form by left-multiplying by $V(\params)$ along with the test function $\ket{\testfunc}$:
\begin{equation}
    \braket{\testfunc | V | \psi} = \braket{\testfunc | V | \phi} + \braket{\testfunc | V G_0 V | \psi}.
\end{equation}
The weak form can then be converted to its discrete form by setting $\psi \to \trial\psi$ and enforcing orthogonality against $\ket{\testfunc_i} = \ket{\psi_i}$ for $i \in [1, \dots, \nbasis]$.%
\footnote{Note that left-multiplying by $V(\params)$ and enforcing orthogonality against $\ket{\testfunc_i} = \ket{\psi_i}$ is different than simply defining $\ket{\testfunc} = V \ket{\psi}$ and enforcing orthogonality against $\ket{\testfunc_i} = V_i \ket{\psi_i}$ because $V(\params)$ depends on $\params$.
Thus, this is indeed a purely Galerkin approach, rather than a Petrov-Galerkin approach.
}
This yields then Equation~\eqref{eq:schwinger-linear}, and
so the coefficients found by making Equation~\eqref{eq:schwinger-vp} stationary are indeed identical to those found via the Galerkin procedure for Equation~\eqref{eq:ls-wave-function}.

Using the emulation of $\psi$, which is calculated by inserting the optimal coefficients obtained from Equation~\eqref{eq:schwinger-linear} into the definition of $\trial\psi$, we can get the associated $K$ through 
%
\begin{align}
\braket{\phi' | K | \phi} & = \braket{\phi' | V | \psi} 
\notag \\
& \approx \braket{\phi' | V | \trial\psi} \notag \\
& = \sum_{ij} \braket{\phi' | V | \psi_i} (W^{-1})_{ij} \braket{\psi_j | V | \phi}.
\end{align}
This Equation is exactly the solution for $K$ found via the LS equation while assuming a finite-rank approximation for $V$:
\begin{equation}
    V^{f} = \sum_{ij} V \ket{\psi_i} \Lambda_{ij} \bra{\psi_j} V,
\end{equation}
where
\begin{equation}
    (\Lambda^{-1})_{ij} = \braket{\psi_i | V | \psi_j}.
\end{equation}
It is known that the SVP yields a $K$ matrix that is equivalent to that found via a finite-rank approximation to $V$~\cite{Takatsuka1981SchwingerKohnRelationship,Takatsuka1981Scattering}, which shows that the Galerkin projection described in this Section is identical to the SVP.

\subsection{Newton Emulators}
\label{sec:newton}

The Newton variational principle (NVP) for the $K$ matrix is given by~\cite{Melendez:2021lyq,newton2002scattering}
\begin{equation} \label{eq:nvp}
\begin{split}
    \mathcal{K}[\trial K] &= V + V G_0 \trial K + \trial K G_0 V \\
    & \quad - \trial K G_0 \trial K + \trial K G_0 V G_0 \trial K,
\end{split}
\end{equation}
where $\trial K$ is a trial matrix.
If desired, one could instead emulate $T^{(\pm)}$ by imposing the associated boundary conditions on $G_0$.
Here it is assumed that we have chosen an on-shell energy $E$, which will remain implicit throughout.
A separate emulator can be constructed for each choice of $E$.
The functional Equation~\eqref{eq:nvp} is stationary about exact solutions of the LS equation, \ie, $\mathcal{K}[K+\delta K] = K + (\delta K)^2$.
If we write the trial matrix as a linear combination of exact snapshots
\begin{equation} \label{eq:TritzTrial}
    \trial K = \sum_{i=1}^{\nbasis}\coeff_i K_i,
\end{equation}
then we can construct an emulator of the $K$ matrix in the spirit of the RBM\@.

Unlike some of the VPs discussed so far, the NVP is written here in operator form, without yet projecting it into a basis.
This gives us the freedom to assert that any component $\braket{\phi' | \mathcal{K} | \phi}$ constructed from Equation~\eqref{eq:nvp} is stationary, which yields an emulator for $\braket{\phi' | K | \phi}$.
For example, one could choose $\ket{\phi}$ to be a plane-wave partial-wave basis $\ket{k\ell m}$ with momentum $k$ and angular momentum quanta $(l,m)$, or one could keep the angular dependence explicit via $\ket{\phi}=\ket{\mathbf{k}}$ in a single-particle basis.
Coupled channels could be emulated by choosing the angular momentum quanta differently between $\ket{\phi'}$ and $\ket{\phi}$.
\new{In fact, the NVP does not even require $\ket{\phi}$ and $\ket{\phi'}$ to be free-space states; one can impose stationarity of Equation~\eqref{eq:nvp} between any two states due to its operator form.}
Note that the \emph{independent} coefficients $\coeffs$ are found for each choice of $\ket{\phi'}$ and $\ket{\phi}$.
Thus, in the case of coupled partial waves, for example, each channel is emulated independently.
To compute phase shifts, we must emulate $K$ at the on-shell energy $E=q^2/2\mu$ and thus, $k = k' = q$ for $\ket{\phi}=\ket{k\ell m}$ and $\bra{\phi'}=\bra{k'\ell'm'}$.

Expressed in the chosen basis, simplifying the functional Equation~\eqref{eq:nvp} after inserting Equation~\eqref{eq:TritzTrial} yields~\cite{Melendez:2021lyq}
\begin{align} \label{eq:LS_identity_beta}
    \braket{\phi' | \mathcal{K}(\params, \coeffs) | \phi} 
    = &\braket{\phi' | V(\params) | \phi} 
    + \coeffs^\trans \vec{m}(\params) 
    \notag \\
    & - \frac{1}{2} \coeffs^\trans  M(\params) \coeffs,
\end{align}
with
\begin{subequations} \label{eq:nvp-reduced-mat-and-vec}
\begin{align} \label{eq:m_vec}
    m_i(\params) & = \braket{\phi' | [K_i G_0 V(\params) + V(\params) G_0 K_i] | \phi},\\
    M_{ij}(\params) & = \braket{\phi' |
                       [K_i G_0 K_j - K_i G_0 V(\params) G_0 K_j] | \phi} \notag \\
                       & \quad\null + (i \leftrightarrow j) .
    \label{eq:M_mat}
\end{align}
\end{subequations}
If the potential $V(\params)$ has an affine parameter dependence, 
$\vec{m}$ and $M$
can be efficiently constructed by linear combinations of matrices pre-computed during the emulator's offline stage, resulting
in substantial improvements in CPU time, \eg, for chiral interactions.

By imposing the stationary condition $\partial \mathcal{K} /\partial \coeffs = 0$, one then finds
$\coeffsopt(\params)$ such that $M \coeffsopt = \vec{m}$.
Given that the optimal $\coeffsopt(\params)$ yields a trial matrix Equation~\eqref{eq:TritzTrial} with an error $\delta K$, one can insert $\coeffsopt$ in Equation~\eqref{eq:LS_identity_beta} to obtain an error $(\delta K)^2$.
The resulting emulator $\mathcal{K}_\star(\params) \equiv \mathcal{K}(\params, \coeffsopt)$ is then~\cite{Melendez:2021lyq}
\begin{equation}\label{eq:nvp-emulator}
    \braket{\phi' | K  | \phi} \approx
    \braket{\phi' | \mathcal{K} | \phi} =
    \braket{\phi' | V | \phi} + \frac{1}{2} \vec{m}^\trans M^{-1} \vec{m}.
\end{equation}
%

Reference~\cite{Melendez:2021lyq} studied several applications of the emulator equation~\eqref{eq:nvp-emulator} to short-range potentials with and without the Coulomb interaction and partial-wave coupling. 
They demonstrated that the NVP emulator has remarkable extrapolation capabilities \new{(see Figure~2 in Reference~\cite{Melendez:2021lyq})} and can quickly reproduce high-fidelity calculations of neutron-proton cross sections based on modern chiral interactions with negligible error.

\new{
We now repeat the derivation for the NVP emulator, but instead from the perspective of a Galerkin projection.
Here, we will focus on the case where $\bra{\phi'} = \bra{\phi}$.
We start with the LS equation
\begin{equation} \label{eq:LS}
    K = V + V G_0 K, 
\end{equation}
which, in this context, constitutes the strong form of the integral equation.
Although Equation~\eqref{eq:LS} is written in terms of abstract operators, it can be turned into a vector equation in a specific representation after right-multiplying by $\ket{\phi}$.
To derive the weak form we left-multiply by $G_0$ and a test function $\bra{\testfunc}$:
%
\begin{align}
    \braket{\testfunc | G_0 K -  G_0 V G_0 K | \phi} & = \braket{\testfunc | G_0 V | \phi} .
\end{align}
%
The trial function in this case is $K\ket{\phi}$, which can be expanded in a discrete (snapshot) basis using Equation~\eqref{eq:TritzTrial}.
We further employ the Galerkin prescription, where the test basis is equivalent to the trial basis, making $\bra{\testfunc_i} = \bra{\phi} K_i$.
With these assumptions, the reduced weak form becomes
}
%
\begin{align}
    M\coeffsopt = \vec{m},
\end{align}
with $M$ and $\vec{m}$ defined in Equation~\eqref{eq:nvp-reduced-mat-and-vec}, again with $\bra{\phi'} = \bra{\phi}$.
Thus, we find the same $\coeffsopt$ using either the NVP or the Galerkin projection.

Given the optimal coefficients $\coeffsopt$, the emulator can be derived by substituting $\trial K$ into the right-hand side of Equation~\eqref{eq:LS}:\new{
\begin{align}
    \braket{\phi | K |\phi} & \approx \braket{\phi | V |\phi} + \braket{\phi | V G_0 \trial K |\phi} \notag\\
    & = \braket{\phi | V |\phi} + \frac{1}{2} \vec{m}^\trans M^{-1} \vec{m},
\end{align}
which is equivalent to Equation~\eqref{eq:nvp-emulator} under the assumption that $\bra{\phi'} = \bra{\phi}$.}
Therefore, both the NVP and Galerkin projection lead to identical emulators for the $K$ matrix.

\subsection{Origin emulators} \label{sec:scattering-origin} 

The scattering emulators discussed so far are 
best known as VPs but are equivalent to various types of Galerkin projections of the Schrödinger or LS equation (see Table~\ref{tab:scattering-vps-galerkin}).
However, other types of emulators can be constructed via Galerkin projections, even if they do not necessarily correspond to any well-known VP\@.

Starting from the Schrödinger equation---a second-order differential equation---we must impose two boundary conditions.
The first is that $r\psi(r)$ vanishes at $r=0$; this constraint has been automatically satisfied by our choice of trial bases in all VPs considered above.
But the second constraint is yet to be chosen.
In the KVP, for example, the second constraint was obtained via the normalization of $r\psi(r)$ as $r\to\infty$, which, \new{in the constrained KVP,} led to a normalization constraint for the coefficients $\coeff_i$.
Because the Schrödinger equation is linear and homogeneous, this normalization condition is equivalent to imposing a constraint on the derivative of $r\psi(r)$, \eg, evaluated at the origin.

Thus, an alternative weak form for the Schrödinger equation could be constructed using only constraints at the origin.
Let us construct a coordinate-space emulator with $(r\psi)'(0) = 1$.
Starting from the generic weak form~\eqref{eq:weak_differential}, we obtain
\begin{align}
    \braket{\zeta | H - E | \psi} + \left.\bar\zeta \left[(r\psi)' - 1\right]\right|_{r=0} = 0,
\end{align}
where $\zeta$ and $\bar\zeta$ are the (independent) test functions in the domain and on the boundary, respectively, and the boundary condition is only evaluated at the origin.
Here, we can make the Galerkin choice of (domain) test functions, where $\bra{\zeta} = \bra{\psi}$, but make a Petrov-Galerkin choice for the boundary, with $\bar\zeta(0) = 1$.

Thus, the discretized weak form, from which our emulator equations follow, is given by
\begin{align}
    \braket{\psi_i | H - E | \psi_j}\coeff_j + \sum_j \coeff_j - 1 = 0,
\end{align}
where we have assumed that the trial basis is constructed such that each snapshot satisfies $(r\psi_j)'(0) = 1$.

\begin{figure}
    \centering
    \includegraphics{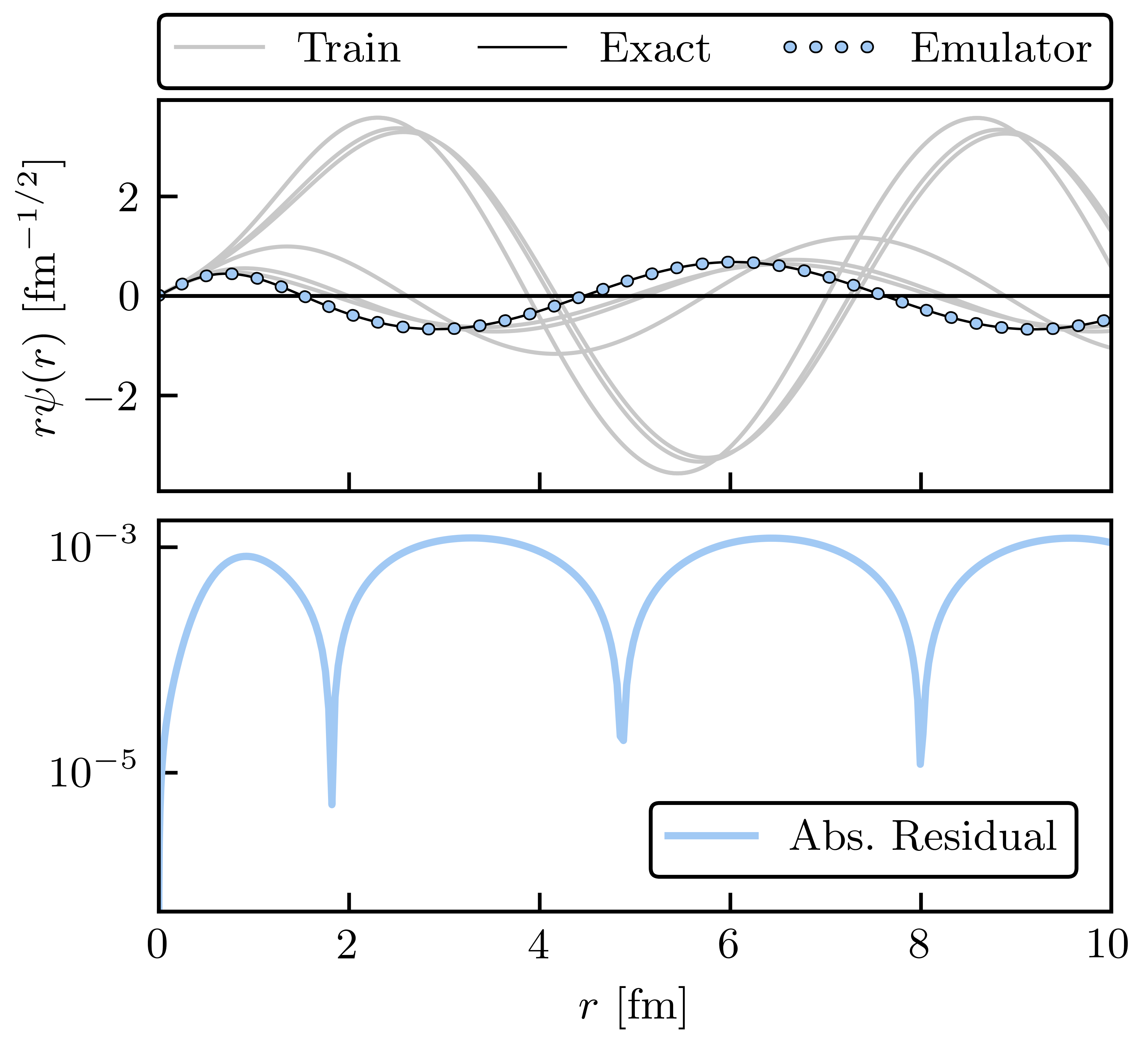}
    \caption{Results for the scattering emulator with origin boundary conditions.
    Six basis functions are shown as gray lines, and the exact wave function as a black line.
    Each of the basis functions and the emulated wave function satisfy the constraints at the origin.}
    \label{fig:wave-functions-origin}
\end{figure}

As an example, we show the output of such an emulator in Figure~\ref{fig:wave-functions-origin}.
Here, the potential is given by a sum of two Gaussians,
\begin{equation}
    V(r, \params) = \theta_1 \exp(-\kappa_1 r^2) + \theta_2 \exp(-\kappa_2 r^2),
\end{equation}
with $\kappa_1 = 0.5$ and $\kappa_2 = 1$. 
\new{The parameters to be varied are $\params = \{\theta_1, \theta_2\}$.}
The six training and one validation parameters are selected randomly from a uniform distribution in the range of $[-5, 5]$\,MeV.\@
To obtain the snapshots, the partial-wave decomposed radial Schrödinger equation can be expressed as the system of coupled first-order differential equations,
\begin{equation}
\begin{pmatrix}
    y'_0(r)\\
    y'_1(r)
\end{pmatrix} = 
\begin{pmatrix}
    \psi'(r) \\
    \frac{\ell(\ell+1)}{r^2} + 2\mu \left[V(r; \bm{\theta}_i)  - E \right] \psi(r)
\end{pmatrix},
\end{equation}
and numerically solved with Runge-Kutta methods. 
For more details on solving the radial Schrödinger equation and matching the solutions to the asymptotic boundary condition~\eqref{eq:psi-normalization}, see, \eg, Reference~\cite{thompson2009}.
As we can see from Figure~\ref{fig:wave-functions-origin}, each training and emulated wave function has matching boundary conditions at the origin, and the discrepancy from the true wave function is less than $10^{-3}$.

\subsection{General Kohn Variational Principle} \label{sec:kohn-general}

The KVP functional given in Equation~\eqref{eq:kohn-psi} can be extended to include arbitrary boundary conditions~\cite{Drischler:2021qoy,Lucchese:1989zz}. 
For simplicity, we demonstrate the general KVP by using a single-channel example in coordinate space, but this work also holds for two-body scattering in momentum space (see Reference~\cite{KVP_vs_NVP:2022}).
Let us consider short-range potentials $V(\params)$ in coordinate space and partial-wave decomposed into an uncoupled channel with angular momentum $\ell$.
%
%
%
The radial wave functions will be the free-space solutions%
\footnote{We follow the conventions for scattering matrices in References~\cite{taylor2006scattering,Morrison:2007}.}
of the general form
\new{\begin{align} \label{eq:free-sol-coordinate}
\phi_{\ell,E}(r) =  \bar{\phi}_{\ell,E}^{\text{(0)}}(r) + L_{\ell,E} \, \bar{\phi}_{\ell,E}^{\text{(1)}}(r) \, ,
\end{align}}
where 
\begin{equation}\label{eq:freeSolMatch}
\begin{pmatrix}
    \bar{\phi}_{\ell,E}^{\text{(0)}}(r)\\
    \bar{\phi}_{\ell,E}^{\text{(1)}}(r)
\end{pmatrix} =
\mathcal{N}^{-1}
\begin{pmatrix}
    u_{00} & u_{01}\\
    u_{10} & u_{11}\\
\end{pmatrix}
\begin{pmatrix}
     j_\ell(qr) \\
    \eta_\ell(qr)
\end{pmatrix} \,,
\end{equation}
with $q = \sqrt{2\mu E}$ and an arbitrary normalization constant $\mathcal{N} \neq 0$. 
Here, $L_{\ell, E}$ is a generic scattering matrix that is determined by the boundary condition, as parametrized by the nonsingular matrix $\umatrix$.

We now define $\genkvp$ as a general functional for the generic $L$-matrix in Equation~\eqref{eq:free-sol-coordinate}~\cite{Lucchese:1989zz,Drischler:2021qoy,KVP_vs_NVP:2022},
\begin{equation} \label{eq:kvp-general}
    \mathcal{L}[\trial\psi] = L_{\ell, E} + \frac{\mathcal{N}}{\mathrm{det} \, \umatrix} \braket{\trial \psi_{\umatrix} | H - E | \trial \psi_{\umatrix}}.
\end{equation}
With Equation~\eqref{eq:kohn-psi}, one can follow the process described in Section~\ref{sec:kohn-lagrange} to emulate any asympototic boundary condition.
Obtaining an emulator prediction for different boundary conditions does not mean that Equation~\eqref{eq:kvp-general} has to be solve multiple times. 
In fact, it only needs to be solved once and each term in the functional \emph{rescaled} using the relations derived in Reference~\cite{Drischler:2021qoy}:
\begin{align}
    \dU^{(\umatrix')}
    &=
    C^{'-1}(L_i) \, C^{'-1} (L_j) \frac{\mathrm{det} \, \umatrix}{\mathrm{det} \, \umatrix'} \dU^{(\umatrix)},  \label{eq:kohn_rescaling_dU}
    \\
    C'(L) 
    &= \frac{\mathrm{det} \, \umatrix}{\mathrm{det} \, \umatrix'} \frac{u'_{11} - u'_{10} K(L)}{u_{11} - u_{10} K(L)}. \label{eq:kohn_c_coeff}
\end{align}
The non-primed terms refer to the initial state and primed terms refer to the final state (explained below). 
The snapshots used to train the emulator in the offline stage are transformed using the Möbius (or linear fractional) transform
\begin{align} \label{eq:mobius_transform}
    L'(L) =  \frac{-u'_{01}+u'_{00} K(L)}{u'_{11} -u'_{10}K(L)}.
\end{align}

Let us consider solving Equation~\eqref{eq:kvp-general} using the $K$-matrix boundary condition, but then wanting a prediction for the $T$-matrix. We would first 
rescale $\dU$ using Eq~\eqref{eq:kohn_rescaling_dU}. Here, $\umatrix$ and $\umatrix'$ would correspond to $\umatrix_K$ and $\umatrix_T$, respectively, given by
\begin{equation} \label{eq:u_matrices}
    \umatrix_K = 
    \begin{bmatrix}
        1 & 0 \\
        0 & 1
    \end{bmatrix},
    \quad
    \umatrix_T =
    \begin{bmatrix}
        1 & 0 \\
        i & 1
    \end{bmatrix}.
\end{equation}
Once $\dU^{(\umatrix')}$ is calculated and the snapshots transformed from the $K$- to the $T$-matrix according to Equation~\eqref{eq:mobius_transform}, we apply Equation~\eqref{eq:kohn-psi-coeff-solution} to obtain the emulator prediction for the $T$-matrix. One can also inverse transform the new emulated solution back to its $K$-matrix equivalent by using
\begin{align}
    K(L) = \frac{u_{01} + u_{11} L}{u_{00} + u_{10} L} .
\end{align}

Variational principles may not always provide a (unique) stationary approximation, causing the appearance of spurious singularities known as Kohn (or Schwartz) anomalies~\cite{Drischler:2021qoy,Lucchese:1989zz,nesbet1980variational}, which can render applications of VPs ineffective; especially for sampling of a model's parameter space.\footnote{\new{These anomalies are not restricted to the KVP but also appear in other VPs such as the NVP and SVP~\cite{ADHIKARI1991435}.}}
The appearance of these anomalies depends on the parameters $\params$ used to train the emulator in the offline stage, the scattering energy, and the evaluation set used in the online stage.
However, Reference~\cite{Drischler:2021qoy} demonstrated that a KVP-based emulator that simultaneously emulates an array of KVPs with different boundary conditions can be used to systematically detect and remove these anomalies. 
\new{The anomalies can be detected by assessing the (relative) consistency of the different emulated results for, \eg, the scattering $S$ matrix. 
The results that do not pass the consistency check are discarded and the remaining ones averaged to obtain an anomaly-free estimate of the $S$ matrix (or any other matrix). 
If all possible consistency checks fail, one can change the basis size of the trial wave function iteratively, which typically shifts the locations of the Kohn anomalies in the parameter space in each iteration.}
The basic idea for removing Kohn anomalies is general and can be applied to other emulators, including the NVP-based emulator discussed in Section~\ref{sec:newton}, as long as multiple scattering boundary conditions can be emulated independently and efficiently.
\new{Alternatively, one might also consider comparing the consistency of emulated results obtained from different VPs such as the ones summarized in Table~\ref{tab:scattering-vps-galerkin}.}

\subsubsection{Generalization to coupled systems} \label{sec:kohn-coupled}


Following Reference~\cite{KVP_vs_NVP:2022}, let us now extend the generalized KVP in Section~\ref{sec:kohn-general} to coupled systems, which could be coupled partial-wave or reaction channels. 
The stationary approximation to the high-fidelity $L$-matrix then reads
\begin{equation}\label{eq:kvp_reduced_coupled}
    \mathcal{L}^{ss'} = \beta_i L_i^{ss'} + \frac{1}{2} \beta_i \dU_{ij}^{ss'}\beta_j,
\end{equation}
where
\new{
\begin{align}
    \dU_{ij}^{ss'} & \equiv \frac{1}{\mathrm{det} \, \umatrix} 
    \big[
    \braket{\psi_i^{st} | [H(\params) - E]^{tt'} | \psi_j^{t's'}} \notag \\
    & \qquad \qquad \, + (i \leftrightarrow j) \notag
    \big]
    \notag\\
    & = \frac{1}{\mathrm{det} \, \umatrix} 
    \big[
    \braket{\psi_i^{st} | [V(\params) - V_j]^{tt'} | \psi_j^{t's'}} \notag \\
    & \qquad \qquad \, + (i \leftrightarrow j) \notag
    \big], \notag\\
    \label{eq:delta_u_tilde_coupled}
\end{align}
}%
with $s$ and $s'$ corresponding to the entrance and exit channels and $t$ and $t'$ are summed over the available channels.
The uncoupled case is retrieved by replacing $ss' \to \ell$.
Solving for $\coeffs$ now proceeds as in Equation~\eqref{eq:kohn-psi-coeff-solution} but for a specific choice of $ss'$ channels.
Note that the coefficients $\coeffs$ are to be determined \emph{independently} for each $ss'$ pair.

\new{The coefficients are independent because $\mathcal{L}^{ss'}$ is independently stationary for each $ss'$ pair.} 
This becomes apparent when considering how one would solve for the coefficients in the case there are two uncoupled channels, where $ss' \to \ell$ in Equations~\eqref{eq:kvp_reduced_coupled} and~\eqref{eq:delta_u_tilde_coupled}.
Here, each partial wave is completely independent of one another,
and thus each VP and their corresponding coefficients $\coeffs$ are independent of one another across values of $\ell$.
Without loss of generality, let the two channels be labeled as $\ell = 0$ and $\ell = 1$, and let $\coeffs^{(0)}$ and $\coeffs^{(1)}$ denote the independent sets of coefficients found by making each channel's KVP stationary.
Now consider adiabatically turning on a coupling between two partial waves: the coefficients $\coeffs^{(0)}$ and $\coeffs^{(1)}$ should remain nearly fixed to their previously uncoupled values, but now there is a new set of coefficients to determine, which one could label as $\coeffs^{(01)}$.
Thus, even in the coupled case, there are multiple independent sets of coefficients to determine: one for each pair of incoming and outgoing channels.

An alternative way to understand how the $\coeffs$ enter in the coupled case is to instead start with the Schrödinger equation and enforce \mbox{(Petrov-)}Galerkin orthogonalization as in Section~\ref{sec:kohn-lagrange}.
For the diagonal channels, the test functions are chosen to have the same outgoing channel as the trial functions, making the procedure of standard Galerkin form.
But for the off-diagonal channels, the test functions have a different outgoing channel ($s$) than the trial functions ($s'$).
Because the basis of test functions differs from the basis of the trial function, this is instead a Petrov--Galerkin approach.
The linear equations to be solved are exactly what one would obtain from enforcing stationarity in Equation~\eqref{eq:kvp_reduced_coupled} for each $ss'$ independently. \new{See Reference~\cite{KVP_vs_NVP:2022} for more information on coupled channel emulation}.

\subsubsection{Generalizations to higher-body systems} \label{sec:higher-body-continuum}

The variational emulators for two-body scattering described so far can be generalized to higher-body scattering. 
In fact, the KVP, as a powerful method for solving scattering problems, has been applied in developing high-fidelity solvers (as opposed to a KVP-based emulator) for studying three- and four-nucleon systems (\eg, nucleon-deuteron elastic scattering below and above the deuteron break-up threshold)~\cite{Marcucci:2019hml, Kievsky:2008es}.\footnote{The other high-fidelity solvers in this context solve problems in momentum or coordinate space based on the Faddeev formalism~\cite{Gloeckle:1995jg, Deltuva:2012kt, Lazauskas:2019rfb}.}
It is then natural to combine the KVP with the variational emulation strategy to develop fast \& accurate emulators beyond just two-body scattering.  

Here, we follow Reference~\cite{Zhang:2021jmi}, which developed KVP-based emulators for three-body systems. 
We focus on systems of three identical spinless bosons, particularly the elastic scattering between boson and two-boson bound state in the channel without any relative angular momenta and below the bound state's break-up threshold. 
The corresponding scattering $S$-matrix can be estimated via a variational functional that resembles 
Equation~\eqref{eq:kvp-general} in the two-body case:
\begin{equation}\label{eq:three-body-elastic-scattering-KVP-functional}  
 \mathcal{S}[\trial\psi] = S - \frac{i}{3\mathcal{N}^2}  \braket{\trial\psi | [H - E] | \trial\psi} . 
\end{equation}
Here, $S$ is the $S$-matrix associated with the trial three-body wave function $|\trial\psi\rangle$, $H$ and $E$ the full Hamiltonian and energy, respectively. 
The trial wave function has the following asymptotic behavior~\cite{Zhang:2021jmi}:
\begin{equation} \label{eq:WFbelowbreakupAsym}
\langle {R}_1, {r}_1|\trial\psi\rangle 
      \overset{R_1\to \infty}{\longrightarrow} \frac{\mathcal{N}}{\sqrt{v} } \frac{u_B({r}_1) }{ r_1 R_1} 
     \bigl(-e^{-i P R_1} + S\, e^{i P R_1} \bigr)  ,  
\end{equation}
with $R_1, r_1$ as one of three different Jacobi coordinate sets; $v$ and $P$ as the relative velocity and momentum between the scattering particles, $\mathcal{N}$ the normalization constant that also appeared in Equation~\eqref{eq:three-body-elastic-scattering-KVP-functional}, and $u_B(r_1)$ the radial wave function of the two-body bound state. 

The emulation procedure is generally similar to those for two-body emulations.
We first collect high-fidelity calculations of $\ket{\trial\psi_i}$ at various points in the Hamiltonian's parameter space during the offline (\ie, training) stage and then use these snapshots as the basis to construct the trial solution [see Equation~\eqref{eq:trial_general_subeq}] to be used in the variational functional during the online emulation stage.
A similar set of the low-dimensional linear equations as in Equations~\eqref{eq:kohn-psi-coeff-solution} can be derived to fix the weights $\beta_i$.
The variational functional with these inputs produce accurate results for the $S$ matrix at the emulation points~\cite{Zhang:2021jmi}. 
This is all straightforward if we vary only the three-body interactions in $H(\params)$ when exploring its parameter space.
If the two-body interactions are also changed, the two-body bound states of those snapshots are different among themselves and therefore the trial wave functions based on Equation~\eqref{eq:trial_general_subeq} fails to satisfy the asymptotic behavior described in Equation~\eqref{eq:WFbelowbreakupAsym}.
In Reference~\cite{Zhang:2021jmi}, proper modifications were applied to the constructions of the trial wave functions to satisfy the asymptotic condition.
The resulting emulator is again a low-dimensional equation system, but the projected $ \subspace{H}(\params)$ matrices and the $\dU$ matrices lose the affine structure as needed for fast emulation [see Equation~\eqref{eq:Htilde_affine}].
To mitigate this issue, the GP emulation method was employed to interpolate and extrapolate the $\dU$'s matrix elements in the parameter space (note that this dependence is much smoother than the parameter dependence of the observables).
Other hyperreduction approaches~\cite{Benner20201} could also be explored in this context. 

The results in Reference~\cite{Zhang:2021jmi} are encouraging: the time cost for emulating three-boson scattering is on the order of milliseconds (on a laptop), while the emulation's relative errors vary from $10^{-13}$ to $10^{-4}$ depending on the case.
It is straightforward to generalize it to elastic scattering above the break-up threshold, but more studies need to be done for emulating the break-up processes and even higher-body systems. 
Of course, the Fermi statistics, spin and isospin degrees of freedom, and partial waves beyond the s-wave need to be included to realize emulation for realistic three and higher-body scatterings (\eg for three-nucleon systems). 
\subsection{A scattering example} \label{sec:scattering-example}

We have covered the reduced-order models that can be constructed from the Kohn, Schwinger, and Newton VPs, and now we put them into action.
This example is given in the context of a rank-$n$ separable potential where simple analytic forms are available for the snapshots.
This provides a sandbox to explore many aspects of the RBM for quantum scattering without the complicating details of more realistic systems.
All of the source code that generates the results shown here is available to explore on the companion website~\cite{companionwebsite}.

Separable potentials lead to simple formulas for the $K$ matrix and the scattering wave function~\cite{Tabakin:1969mr}.
A rank-$n$ separable potential in momentum space is given by
\begin{align} \label{eq:separable-potential}
    V_{\ell} = \sum_{ij}^n \ket{v_i^\ell} \Lambda_{ij} \bra{v_j^\ell} ,
\end{align}
where $\Lambda_{ij} = \Lambda_{ji}$ are the coefficients of the potential that will be varied during emulation.
For simplicity, we consider here only $s$-wave scattering (\ie, $\ell = 0$). 
The potential~\eqref{eq:separable-potential} leads to an affine structure that lends itself to the offline-online decomposition discussed in Section~\ref{sec:model-reduction}.
From the potential~\eqref{eq:separable-potential}, simple expressions for $K$ and $\psi$ can be derived~\cite{PhysRevC.55.1650}.
For instance, the $K$ matrix is given in operator form by
\begin{align}
    K = \sum_{ijk}^n \ket{v_i} \Lambda_{ij}[\identity - J \Lambda]^{-1}_{jk} \bra{v_k} ,
\end{align}
with the identity matrix $\identity$ and the matrix
\begin{align}
    J_{ij} \equiv \braket{v_i | G_0 | v_j},
\end{align}
where the Green's function $G_0$ implicitly depends on the on-shell energy $E$.
Thus, it follows that $K$ is separable if $V$ is separable.

We choose to study the Yamaguchi potential~\cite{Gobel:2019jba}
\begin{align}
    \braket{p|v_i^\ell} \equiv v_i^\ell(p) = \frac{p^{\ell}}{(p^2 + b_i^2)^{\ell+1}} \label{eq:yamaguchi}
\end{align}
%
with $\ell=0$ and assume a rank-2 potential with $b_i = [2, 4]$\,fm$^{-1}$ and $2\mu = 1$.
\new{In this case,
\begin{align}
    J_{ij} = \frac{\pi}{2} \frac{(q^2 - b_i b_j)}{(b_i + b_j)(q^2 + b_i^2)(q^2 + b_j^2)},
\end{align}
which permits all phase shifts, wave functions, and reduced-order matrices (\eg, $\dU$) to be evaluated analytically.}
The training parameters $\{\Lambda_{00}, \Lambda_{01}, \Lambda_{11}\}$ are sampled randomly from a uniform distribution in $[-50, 50]$\,MeV.
The companion website~\cite{companionwebsite} provides the following Python classes that implement the scattering emulators: \{\texttt{Newton}, \texttt{Schwinger}, \texttt{Kohn}, \texttt{UnconstrainedKohn}\}\-\texttt{Emulator}.

\begin{figure}[tb]
\includegraphics[]{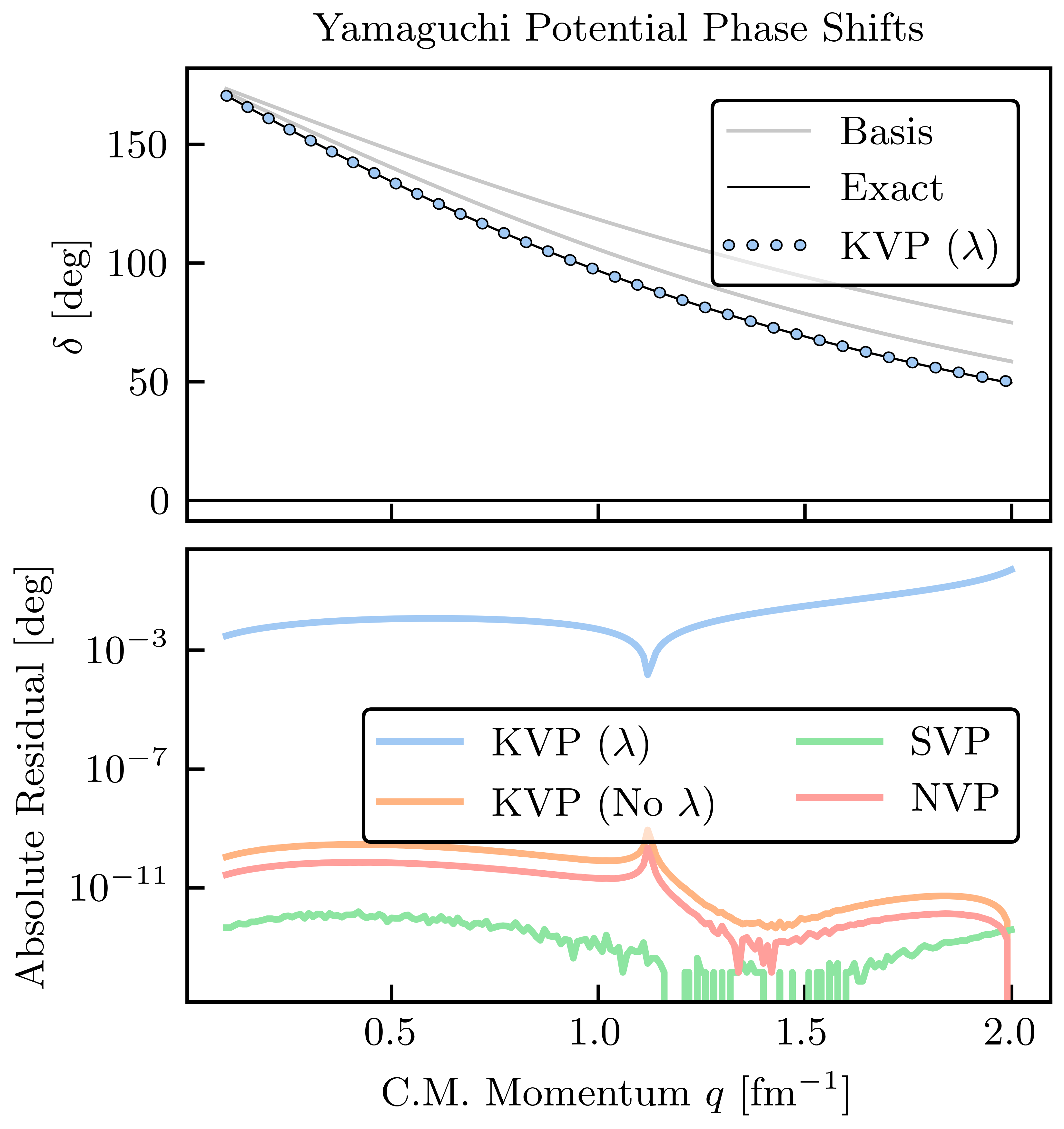}
\caption{
    \new{Phase shifts (top panel) and absolute residuals (bottom panel) for the Yamaguchi potential~\eqref{eq:yamaguchi} for each scattering emulator discussed above. 
    The solid black lines represent the high-fidelity solution and the dots represent the emulator results.
    The emulators are so accurate that they are indistinguishable unless looking at residuals.
    The training set is given by the 
    two gray lines.}
    } \label{fig:yamaguchi_phase_shifts}
\end{figure}

Figure~\ref{fig:yamaguchi_phase_shifts} shows the phase shifts and the absolute residuals for the Yamaguchi potential~\eqref{eq:yamaguchi} for emulators constructed with $\nbasis=2$ training points. 
The top panel depicts the high-fidelity solution (solid black curve) and the emulator results (dots).
\new{Here, only the constrained KVP is shown because the others would be indistinguishable.}
In gray we show the basis states used to train the emulator in the offline stage. 
The bottom panel shows the absolute residuals for each of the emulators. 
\new{We can see that all but the constrained KVP are extremely accurate, with the residuals mostly governed by the choice of nugget used to regularize the matrix inversion (see also Section~\ref{sec:eigen-emulators_variational}).
For the constrained KVP, we see that the loss of a degree of freedom to implement the constraint significantly impacts its predictive power given that we only have two basis states, although it is still quite accurate in this case.
Increasing the basis to $\nbasis = 5$ yields predictions that are accurate to $10^{-13}$ degrees, or better, for all emulators.}

\begin{figure}[tb]
\includegraphics[]{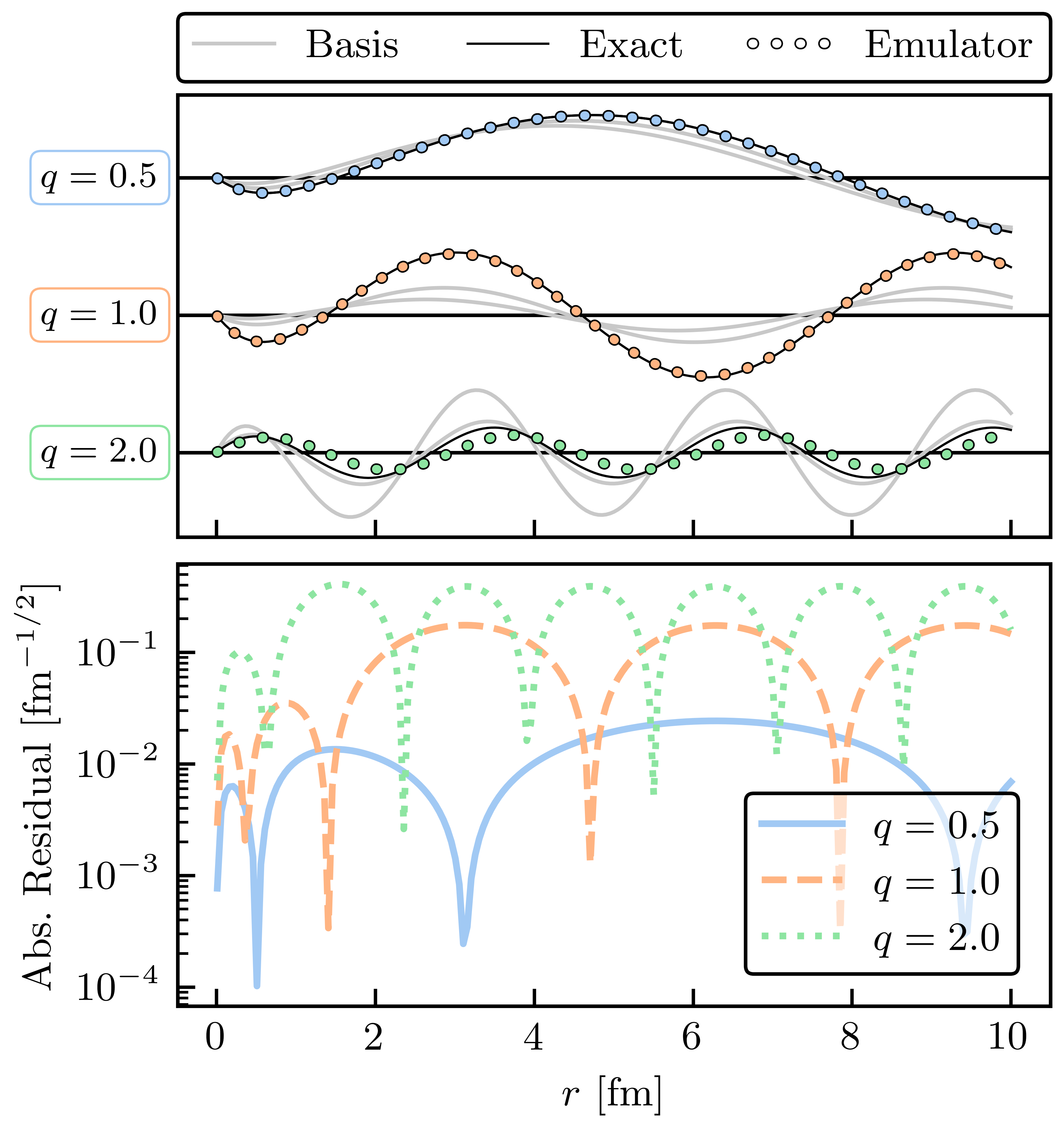}
\caption{
    \new{Wave functions (top panel) and absolute residuals (bottom panel) for the Yamaguchi potential~\eqref{eq:yamaguchi} using the constrained KVP.\@ 
    The top panel legend description is similar to Figure~\ref{fig:yamaguchi_phase_shifts}, but for three different values of $q$ in units of fm$^{-1}$. 
    The bottom panel shows the relative residuals of the three values previously mentioned.}
    } \label{fig:yamaguchi_wave_functions}
\end{figure}

Figure~\ref{fig:yamaguchi_wave_functions} shows the high-fidelity (solid black line), emulated (dots), and basis (solid gray line) wave functions for three values of $q$ with their absolute residuals using the constrained KVP constructed with $\nbasis=5$ training points. 
The emulator reproduces the high-fidelity solution at all three values of $q$, with $q = 2.0 \, \text{fm}^{-1}$ having the smallest residual.
\new{The sensitivity of the emulator accuracy as $\nbasis$ is varied can be readily studied using the Python code provided on the companion website~\cite{companionwebsite}. 
An example of how the accuracy is affected when varying $\nbasis$ is also given in Reference~\cite{KVP_vs_NVP:2022}.}

\new{While all emulators described in this Section are applicable to scattering problems in general, their efficacy will depend in practice on various factors, such as their computational complexity and the potential to be emulated.
The constrained KVP has the advantage that terms constant in $\params$, such as the (long-range) Coulomb potential, cancel in the computation of Equation~\eqref{eq:delta_u_tilde} but it loses one degree of freedom due to the normalization constraint of the coefficients.
On the other hand, both the NVP and SVP involve the computation of Green's functions, which makes them computationally more complex than the KVP---especially the SVP
since it also depends quadratically on the potential.
}

\section{Summary \& Outlook}
\label{sec:conclusion}

We have presented a pedagogical introduction to projection-based, reduced-order emulators and general MOR concepts suitable for a wide range of applications in low-energy nuclear physics.
Emulators are fast surrogate models capable of reliably approximating high-fidelity models due to their reduced content of superfluous information.
By making practical otherwise impractical calculations, they can open the door to the various techniques 
and applications central to the overall theme of this Frontiers Research Topic~\cite{ResearchTopicUQ}, such as Bayesian parameter estimation for UQ, experimental design, and many more.

In particular, we have discussed variational and Galerkin methods combined with snapshot-based trial (or test) functions as the foundation for constructing fast \& accurate emulators.
These emulators enable repeated bound state and scattering calculations, \eg, for sampling a model's parameter space when high-fidelity calculations are computationally expensive or prohibitively slow.
A crucial element in this emulator workflow, as summarized in Figure~\ref{fig:illustration_fom_rom}, is an efficient offline-online decomposition in which the heavy computational lifting is performed only once before the emulator is invoked. 
Chiral Hamiltonians allow for such efficient decompositions due to their affine parameter dependence on the low-energy couplings.
\new{Furthermore, we discussed the high efficacy of projection-based emulators in extrapolating results far from the support of the snapshot data, as opposed to the GPs.
} 


While MOR has already reached maturity in other fields, it is still in its infancy in nuclear physics---although rapidly growing---and there remains much to explore and exploit~\cite{Melendez:2022kid,Bonilla:2022rph,Anderson:2022jhq,Giuliani:2022yna}. 
In the following, we highlight some of the many interesting research avenues for emulator applications in nuclear physics.
All of these avenues can benefit from the rich MOR literature and software tools available (\eg, see References~\cite{Benner_2017aa,Benner2017modelRedApprox,benner2015survey}):
\begin{itemize}
    \item Emulator uncertainties need to be robustly quantified and treated on equal footing with other uncertainties in nuclear physics calculations, such as EFT truncation errors.
    This will be facilitated by the extensive literature on the uncertainties in the RBM~\cite{chen2017RBUQ,Haasdonk2016RBM,HUYNH2007473,Rozza2008}.

    \item \new{The performance} of competing emulators (\eg, the Newton vs Kohn variational principle) is typically highly implementation dependent. 
    Best practices for efficient implementation of nuclear physics emulators should be developed.
    \new{This may include exploiting MOR software libraries from other fields, such as \texttt{pyMOR}~\cite{milk2016pyMOR}, when possible.}

    \item Galerkin emulators are equivalent to variational emulators for bound-state and scattering calculations if the test and trial basis are properly chosen. 
    But Galerkin (and especially Petrov-Galerkin) emulators are more general and exploring their applications 
    in nonlinear problems
    may be fruitful in nuclear physics. 
    \new{Emulator applied to non-linear problems will have challenges in terms of both speed and accuracy: 1) the basis size will, in general, need to be large(r) resulting in lower speed-up factors and longer offline stages;
    2) using hyperreduction methods will lead to additional approximations that worsen the accuracy of the emulator and whose uncertainties need to be quantified.}
    
    \item Many technical aspects should be further explored, such as greedy (or active-learning)~\cite{Sarkar:2021fpz} and SVD-based algorithms for choosing the training points more effectively, hyperreduction methods for non-affine problems, and improved convergence analyses.
    
    \item \new{Scattering emulators could play pivotal roles in connecting reaction models and experiments at new-generation rare-isotope facilities (\eg, the Facility for Rare Isotope Beams).
    In this regard, further studies on incorporating long-range Coulomb interactions and optical potentials beyond two-body systems will be valuable. 
    Furthermore, emulators for time-dependent density functional theories could see extensive applications in interpreting fission measurements. 
    At facilities such as Jefferson Lab and the future Electron-Ion Collider, explorations of nuclear dynamics at much higher energy scales
    should also benefit from emulators.} 
    
    \item The emulator framework can be used to extrapolate observables far away from the support of the training data, such as the discrete energy levels of a many-body system calculated in one phase to those of another, as demonstrated in Reference~\cite{Frame:2017fah}. 
    Using emulators as a resummation tool to increase the convergence radius of series expansions~\cite{Demol:2019yjt} falls into this category as well, and so does using them to extrapolate finite-box simulations of quantum systems across wide ranges of box sizes~\cite{Yapa:2022nnv}.
    Moreover, for general quantum continuum states, emulation in the complex energy plane can enable computing scattering observables with bound-state methods~\cite{XilinZhangLECM2022}. 
    Extrapolation capabilities of emulators should be investigated further.
    
    \item While projection-based emulators have had successes (\eg, see References~\cite{Konig:2019adq, Ekstrom:2019lss, Wesolowski:2021cni}), it is also important to understand their limitations and investigate potential improvements. 
    The synergy between projection-based and machine learning methods~\cite{Boehnlein:2021eym} is a new direction being undertaken in the field of MOR for this purpose (\eg, see Reference~\cite{FRESCA2022114181}). 
    Nuclear physics problems, with and without time dependence, will provide ample opportunities for such explorations. 
    
    \item Emulators run fast, often with a small memory footprint, and can be easily shared.
    These properties make emulators effective interfaces for large expensive calculations, through which  users can access sophisticated physical models at a minimum cost of computational resources and without specialized expertise, creating a more efficient workflow for nuclear science. 
    As such, emulators can become a collaboration tool~\cite{Zhang:2021jmi,Drischler:2022yfb} that can catalyze new direct and indirect connections between different research areas and enable novel studies. 
\end{itemize}
To help foster the exploration of these (and other) research directions in nuclear physics, we have created a companion website~\cite{companionwebsite} containing interactive supplemental material and source code so that the interested reader can experiment with and extend the examples discussed here.

We look forward to seeing more of the MOR methodology implemented as these research directions are being pursued. 
But especially we look forward to the exciting applications of emulators in nuclear physics that are currently beyond our grasp. 

\section*{Conflict of Interest Statement}

The authors declare that the research was conducted in the absence of any commercial or financial relationships that could be construed as a potential conflict of interest.

\section*{Author Contributions}


All authors made original and direct contributions to this article and approved it for publication.

\section*{Funding}

This material is based upon work supported by the U.S. Department of Energy, Office of Science, Office of Nuclear Physics, under the FRIB Theory Alliance award DE-SC0013617. 
This work was supported in part by the National Science Foundation under award numbers PHY-1913069 and PHY-2209442 and the NSF CSSI program under award
number OAC-2004601 (BAND Collaboration~\cite{BAND_Framework}), and the NUCLEI SciDAC Collaboration under U.S. Department of Energy MSU subcontract RC107839-OSU\@.

\section*{Acknowledgments}
We thank the Editors Maria Piarulli, Evgeny Epelbaum, and Christian Forss{\'e}n for the kind invitation to contribute to this Frontiers Research Topic~\cite{ResearchTopicUQ}. 
We are also grateful to our BUQEYE~\cite{BUQEYEweb} and BAND~\cite{BAND_Framework} collaboration colleagues for sharing their invaluable insights with us, and especially to Pablo Giuliani and Daniel Phillips for fruitful discussions.
C.D. thanks Petar Mlinari{\'c} for sharing his deep insights into the software library \texttt{pyMOR}~\cite{milk2016pyMOR} for building MOR applications with the Python. 


\section*{Data Availability Statement}
The codes that generate all the figures in this article, interactive tutorials, and supplemental material can be found in Reference~\cite{companionwebsite}.



\bibliographystyle{apsrev4-1}
\bibliography{bibs/bayesian_refs}

\end{document}